\documentclass[10pt,journal,compsoc]{IEEEtran}
\ifCLASSINFOpdf
   \usepackage[pdftex]{graphicx}
\else
\fi
\usepackage{color}
\usepackage[ruled,linesnumbered]{algorithm2e}
\usepackage{amssymb,amsmath}
\usepackage{amsmath}
\usepackage{mathrsfs}
\usepackage{amsmath,amsthm,amssymb,amsfonts}
\usepackage{multirow}
\usepackage{booktabs}
\usepackage{array}
\usepackage{makecell}
\usepackage{stfloats}
\usepackage{epstopdf}
\usepackage{epsfig}
\makeatletter
\thm@headfont{\sc}
\makeatother

\usepackage{url}
\usepackage{cite}

\usepackage{color}
\usepackage[ruled,linesnumbered]{algorithm2e}
\usepackage{amsmath}
\usepackage{amsthm}
\usepackage{ tipa }

\begin{document}

\title{Subgraph Networks with Application to Structural Feature Space Expansion}

\author{Qi~Xuan,~\IEEEmembership{Member,~IEEE},
        Jinhuan~Wang,
        Minghao~Zhao,
        Junkun~Yuan,
        Chenbo~Fu,
        ~Zhongyuan~Ruan,
        and~Guanrong~Chen,~\IEEEmembership{Fellow,~IEEE}
        %~\IEEEmembership{fellow,~IEEE}
	 % <-this % stops a space
\IEEEcompsocitemizethanks{
% \IEEEcompsocthanksitem This work was submitted on September 5, 2018.
\IEEEcompsocthanksitem This article has been accepted for publication in a future issue of IEEE TKDE, but has not been fully edited. Content may change prior to final publication. Citation information: DOI 10.1109/TKDE.2019.2957755
\IEEEcompsocthanksitem Q. Xuan, J. Wang, and C. Fu are with the Institute of Cyberspace Security, College of Information Engineering, Zhejiang University of Technology, Hangzhou, China (e-mail: xuanqi@zjut.edu.cn; JinhuanWang@zjut.edu.cn; yuanjk@zju.edu.cn; cbfu@zjut.edu.cn).
\IEEEcompsocthanksitem M. Zhao is with the Fuxi AI Lab, NetEase Inc., Hangzhou, China (e-mail: zhaominghao@corp.netease.com).
\IEEEcompsocthanksitem J. Yuan is with the College of Computer Science and Technology, Zhejiang University, Hangzhou, China (e-mail: yuanjk@zju.edu.cn).
\IEEEcompsocthanksitem Z. Ruan is with the College of Computer Science and Technology, Zhejiang University of Technology, Hangzhou, China (e-mail: zyruan@zjut.edu.cn).
\IEEEcompsocthanksitem G. Chen is with the Department of Electronic Engineering,
City University of Hong Kong, Hong Kong SAR, China (e-mail: eegchen@cityu.edu.hk).
\IEEEcompsocthanksitem Corresponding authors: Qi Xuan and Zhongyuan Ruan.
}% }
}

\markboth{IEEE Transactions on Knowledge and Data Engineering}
{Shell \MakeLowercase{\textit{et al.}}: Bare Demo of IEEEtran.cls for IEEE Transactions on Magnetics Journals}

% after the title page when using the two side option.

\IEEEtitleabstractindextext{%
\begin{abstract}
Real-world networks exhibit prominent hierarchical and modular structures, with various subgraphs as building blocks. Most existing studies simply consider distinct subgraphs as motifs and use only their numbers to characterize the underlying network. Although such statistics can be used to describe a network model, or even to design some network algorithms, the role of subgraphs in such applications can be further explored so as to improve the results.
%network modeling and algorithms design.
In this paper, the concept of subgraph network (SGN) is introduced and then applied to network models, with algorithms designed for constructing the 1st-order and 2nd-order SGNs, which can be easily extended to build higher-order ones. Furthermore, these SGNs are used to expand the structural feature space of the underlying network, beneficial for network classification. Numerical experiments demonstrate that the network classification model based on the structural features of the original network together with the 1st-order and 2nd-order SGNs always performs the best as compared to the models based only on one or two of such networks. In other words, the structural features of SGNs can complement that of the original network for better network classification, regardless of the feature extraction method used, such as the handcrafted, network embedding and kernel-based methods.
\end{abstract}

}

\maketitle
\IEEEdisplaynontitleabstractindextext
\IEEEpeerreviewmaketitle
\section{Introduction}\label{section:1}
\IEEEPARstart{M}{any} real-world systems can be naturally represented by networks, such as biological networks~\cite{walter2004visualization, wale2008comparison}, collaboration networks~\cite{nguyen2018learning,xuan2018social}, software networks~\cite{xuan2014focus,mockus2002two}, and social networks~\cite{kim2018social,fu2018link}. Studying the substructure of a network, e.g. its subgraphs, is an efficient way to understand and analyze the network~\cite{ullmann1976algorithm}. In fact, subgraphs are basic structural elements of a network, and distinct sets of subgraphs are usually associated with different types of networks.
%, therefore play an important role in network studies.
In retrospect, as shown in~\cite{balazsi2005topological}, frequent appearance of subgraphs can reveal topological interaction patterns, each of which performs precisely some specialized functions, therefore they can be used to distinguish different communities and various networks.

Up to now, a number of studies on network subgraphs for graph classification have been reported. Ugander et al.~\cite{ugander2013subgraph} treated subgraph frequency as a local property in social network and found that subgraph frequency can indeed provide unique insights for identifying both social structure and graph structure in a large network. Similarly, Vohra~\cite{vohrasubgraph} summarized the network by stacking subgraph frequencies into a vector as a global network property and then classified networks into different groups, where these frequency statistics are implemented through two schemes~\cite{ugander2013subgraph, jha2015path}. Moreover, in the study of biological networks, Grochow et al.~\cite{grochow2007network} proposed a novel algorithm for identifying larger network elements and functional motifs, revealing the clustering properties of motifs through subgraph enumeration and symmetry-breaking.
Without any interaction dependencies between them, these studies simply acquired a sequence of discrete motif entities with feature information such as counting, weight, etc. to describe the underlying network.
Except for subgraph frequency statistics, Benson et al.~\cite{benson2016higher} obtained the corresponding embedding representation through laplacian matrix analysis method. Moreover, in~\cite{wang2017incremental}, an incremental subgraph join feature selection algorithm was designed, which forces graph classifiers to join short-pattern subgraphs so as to generate long-pattern subgraph features. Similarly, Yang et al. ~\cite{yang2018node} proposed the NEST method which combined the motifs and convolutional neural network.

The studies mentioned above try to reveal subgraph-level patterns, which can be considered as network building blocks of particular functions, to capture mesoscopic structure. However, most of them ignored the interaction between these subgraphs, which could be of particular importance to represent the global structure of subgraph-level. In order to address this, we propose a method to establish \emph{Subgraph Networks} (SGNs) of different orders. It can be expected that such SGNs can capture the structural features of different aspects and thus may benefit the follow-up tasks, such as network classification. Briefly, there are three steps to build an SGN from an original network: first, detect subgraphs in the original network; second, choose appropriate subgraphs for a task; third, utilize the chosen subgraphs to build an SGN. Line graph~\cite{harary1960some} thus can be considered as a special SGN, where a link connecting two nodes in the original network is considered as a subgraph, and two subgraphs are connected in the SGN if the corresponding two links share a same terminal node. Clearly more complicated subgraphs can be considered, e.g., three nodes with two links, so as to get a higher-order SGN, as will be further discussed in Sec.~\ref{sec:SGN}. The key point here is that the SGN extracts the representative parts of the original network and then assembles them to reconstruct a new network that preserves the relationship among subgraphs. Our method thus implicitly maintains the higher-order structures under the premise of providing the information of local structures. And, the network structure of SGN can complement the original network and, as a result, the integration of their features will benefit the subsequent structure-based algorithms design and applications.

The main contributions of this work are summarized as follows.

\begin{itemize}
\item A new concept of SGN is introduced, along with algorithms designed for constructing the 1st-order and 2nd-order SGNs from a given network. These algorithms can be easily extended to construct higher-order SGNs.
\item SGN is used to obtain a series of handcrafted structural features which, together with the features automatically extracted by using some advanced network-embedding methods, kernel-based methods and depth model, provide complementary features to those extracted from the original network.
\item SGN is applied to network classification. Experiments on seven groups of networks are carried out, showing that integrating the features obtained from SGN can indeed significantly improve the classification accuracy in most cases, as compared to the same feature extraction and classification methods based only on the original networks.
\end{itemize}

The rest of the paper is organized as follows. In Sec.~\ref{sec:related}, some related work about subgraph and network representation methods are briefly introduced. In Sec.~\ref{sec:SGN}, the definition of SGN is provided and algorithms for constructing the 1st-order and 2nd-order SGNs are designed. In Sec.~\ref{sec:NA}, handcrafted structural features are characterized, for both the original network and SGNs. In Sec.~\ref{sec:Exp}, several automatic feature extraction methods are discussed, whereas SGNs are applied to graph classification for some real-world networks. Finally, Sec.~\ref{sec:Con} concludes the investigation, with a future research outlook.

\section{Related work}\label{sec:related}

In this section, we review the related work of subgraph in graph mining applications and the network representation methods combined with depth models in recent years.

\subsection{Subgraph in Graph Mining}
Recently, subgraphs have been widely applied in the study between entities in networks. For example, in~\cite{thoma2010discriminative, wernicke2006efficient, wernicke2005faster}, different algorithms were designed for detecting network subgraphs. In ~\cite{rotabi2017detecting}, a method for detecting strong ties was proposed using frequent subgraphs in a social network, where it was observed that frequent subgraphs as network structural features could lead to good performances in alleviating the sparse problem for detecting strong ties on the network. By adding time stamp to the topology, temporal frequent subgraphs~\cite{kovanen2011temporal, xuan2015temporal, paranjape2017motifs} were studied for some time-dependent networks, such as social and communication networks, as well as biological and neural networks. Furthermore, subgraphs were also applied to graph clustering. In~\cite{tsourakakis2017scalable}, a graph clustering method was developed based on frequent subgraphs, which can effectively detect communities in a network. Network subgraphs deeply depict the local structural features of the network and have important research value in the application of graph mining.

\subsection{Network Representation}
The combination of subgraph structures and depth models enriches the research methods of the network and brings inspiration to researchers. With the rapid development of deep learning, many graph mining and representation methods have been proposed and tested, with practical applications to, e.g., drug design (through studying chemical compound and proteins data)~\cite{jing2018deep,lane2018comparing} and market analysis (through purchase history)~\cite{cheng2016wide}. Methods like word2vec~\cite{mikolov2013distributed} and doc2vec~\cite{le2014distributed} have shown good performances in natural language processing (NLP), bringing some new insights to the field of graph representation. Inspired by these algorithms, graph2vec~\cite{narayanan2017graph2vec} was proposed, which was shown to be outstanding for graph representation. Among the existing graph mining methods, graph kernel~\cite{vishwanathan2010graph, shervashidze2011weisfeiler, yanardag2015deep} has obtained unanimous praise in recent years, whereas the bottleneck is its high computational cost. As a winner from competitions on a plenty of machine learning problems, convolutional neural network (CNN) has attracted lots of attention, especially in the area of computer vision~\cite{xuan2018automatic}, and it has been reformulated by the new convolution operator for graph structure data~\cite{duvenaud2015convolutional}. It was put forward in~\cite{bruna2013spectral}, referred to as graphconv, the first trial of an analogy of CNN on graphs. Then, graph convolutional network (GCN), designed in~\cite{defferrard2016convolutional} as an extension to the k-localized kernel, resolved the problem of over localization as compared with graphconv. Based on graph neural network (GNN) and capsule, Zhang et al.~\cite{xinyi2018capsule} designed the CapsGNN, which can generate multiple embeddings for each graph to capture network properties from different aspects. This method was extensively tested, and achieve the state-of-the-art results.

Network algorithms benefit from graph embedding by automatically extracting features of arbitrary dimensions. However, such methods still largely rely on the original network, and thus may ignore important hidden structural features. To bridge the gap, we map the original network to different structural spaces, in terms of different SGNs. Different from those existing subgraph-based methods~\cite{yang2018node,benson2016higher,vohrasubgraph} that only enumerate a set of motifs as functional building blocks and then match them in the original network for subsequent representation, our SGN model maps the subgraphs in the original network to the nodes in a higher-order structural space, addressing the connections between the subgraphs. Therefore, it can be considered that SGN provides a general framework to expand the structural feature space, which can be naturally integrated into many graph representation methods to further improve their effectiveness.

\section{Subgraph networks}\label{sec:SGN}
Generally, SGN can be considered as a mapping in network space, which maps the original node-level network to subgraph-level networks.
In this section, SGN is first introduced, followed by algorithms for constructing the 1st-order and 2nd-order SGNs.

\newtheorem{myDef}{Definition}
\begin{myDef}[\textbf{Network}]\label{def:1}
An undirected network is represented by $G(V,E)$, where $V$ and $E\subseteq(V\times{V})$ denote the sets of nodes and links, respectively. The element $(v_i,v_j)$ in $E$ is an unordered pair of nodes $v_i$ and $v_j$, i.e., $(v_i,v_j)=(v_j,v_i)$, for all $i,j=1,2,...,N$, where $N$ is the number of nodes, namely the size of the network.
\end{myDef}

\begin{myDef}[\textbf{Subgraph}]\label{def:2}
Given a network $G(V,E)$, $g_i=(V_i,E_i)$ is a subgraph of $G$, denoted by $g_i\subseteq G$ if and only if $V_i\subseteq V$ and $E_i \subseteq E$. The sequence of subgraphs is denoted as $g = \{g_i\subseteq G|i=1,2,...,n\}$, $n\le N$.
\end{myDef}

\begin{myDef}[\textbf{SGN: Subgraph Network}]\label{def:3}
Given a network $G(V,E)$, the SGN, denoted by $G^*=L(G)$, is a mapping from $G$ to $G^*(V^*,E^*)$, with the sets of nodes and links denoted by $V^*=\{g_j|j = 0,1,...,n\}$ and $E^* \subseteq (V^* \times V^*)$, respectively. Two subgraphs $g_i$ and $g_j$ are connected if they share some common nodes or links in the original network, i.e., $V_i\cap{V_j}\neq\emptyset$. Similarly, the element $(g_i,g_j)$ in $E^*$ is an unordered pair of subgraphs $g_i$ and $g_j$, i.e., $(g_i,g_j)=(g_j,g_i)$, $i=1,2,...,n$ with $n\le N$.
\end{myDef}

According to the definition of SGN, one can see that: (i) subgraph is a part of the original network; (ii) SGN is derived from a higher-order mapping of the original network $G$; (iii) the connecting rule between two subgraphs needs to be clarified. Following the approach of~\cite{agarwal2006higher}, where the problem of graph representation in a domain with higher-order relations is discussed, constructing sets of nodes as $p$-chains, corresponding to points (0-chains), lines (1-chains), triangles (2-chains), etc., here the new framework constructs subgraphs as 1st order, 2nd order, etc. For clarity, three steps in building the new framework are outlined as follows.

\begin{itemize}
\item Detecting subgraphs from the original network. Networks are rich of subgraph structures, with some subgraphs occurring frequently, e.g., motifs~\cite{wernicke2006efficient}. Different kinds of networks may have different local structures, captured by different distributions of various subgraphs.
\item Choosing appropriate subgraphs. Generally, subgraphs should not be too large, since in this case SGN may only contain a very small number of nodes, making the subsequent analysis less meaningful. Moreover, the chosen subgraphs should be connected to each other, i.e., they should share some common part (nodes or links) of the original network, so that higher-order structural information can emerge.
\item Utilizing the subgraphs to build SGN. After extracting enough subgraphs from the original network, connections among them are established following certain rules so as to build SGN. Here, for simplicity, consider two subgraphs. They are connected if and only if they share the same nodes or links from the original network. There certainly can be other connecting rules, leading to totally different SGNs, which will be discussed elsewhere in the future.
\end{itemize}

In this paper, the most fundamental subgraphs, i.e., line and triangle, are chosen as subgraphs, since they are simple and relatively frequently appearing in most networks. Thus, two kinds of SGNs of different orders are constructed as follows.

\subsection{First-Order SGN}
In the case of first-order, a line, or a link, is chosen as a subgraph, based on which SGN is built, denoted by SGN$^{\textbf{(1)}}$. The 1st-order SGN is also known as a line graph, where the nodes are the links in the original network, and two nodes are connected if the corresponding links share a same end node.

\begin{figure}[!t]
	\centering
	\includegraphics[width=1\linewidth]{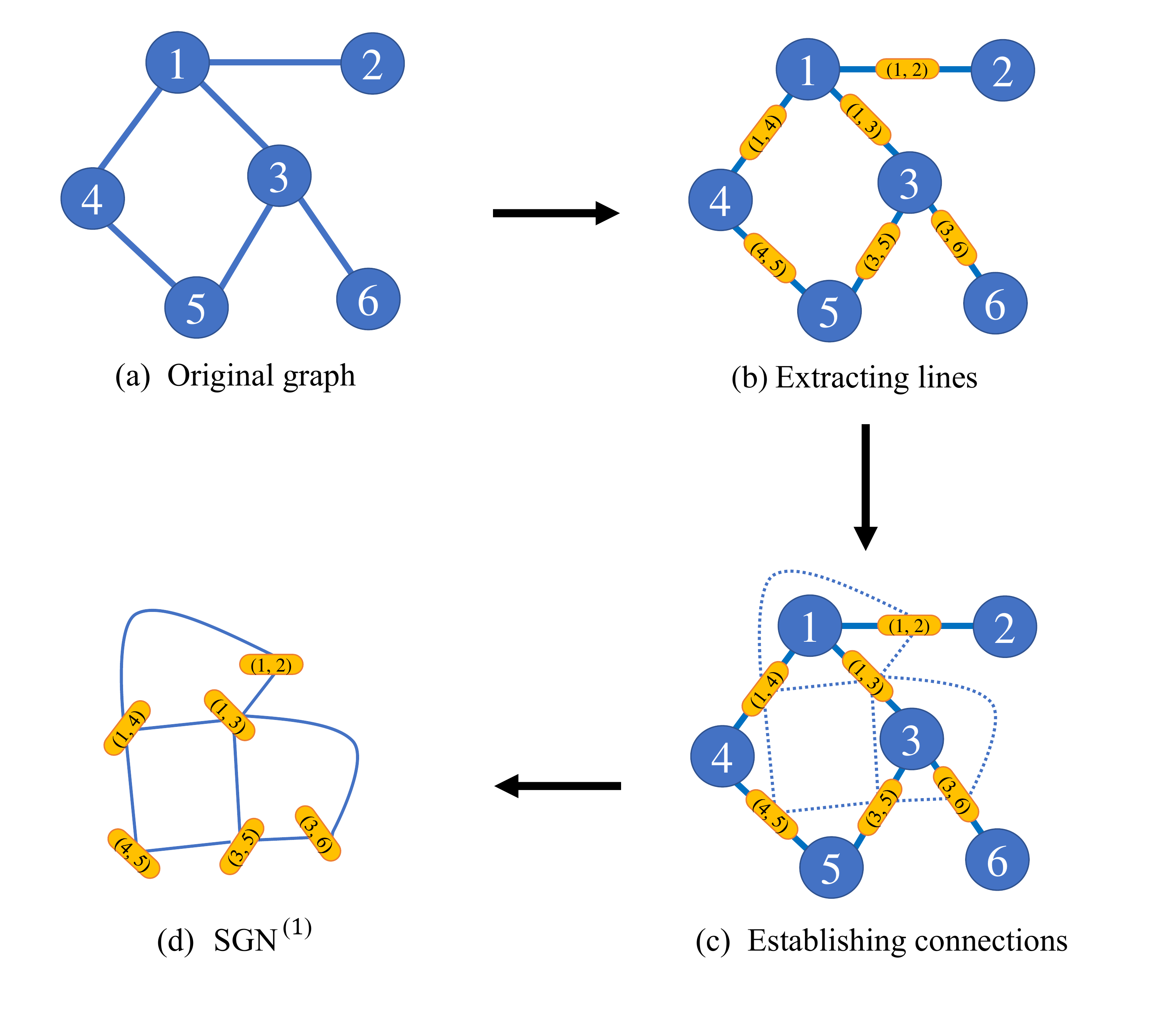}
	\caption{The process of building SGN$^{\textbf{(1)}}$ from a given network: (a) the original network, (b) extracting lines as subgraphs, (c) establishing connections among these lines, and (d) forming SGN$^{\textbf{(1)}}$.}
	\label{fig:line}
\end{figure}

\begin{algorithm}[!t]
\caption{\textbf{Constructing SGN$^{\textbf{(1)}}$.}}
\label{alg:1}
\KwIn{A network $G$($V$,$E$) with node set $V$ and link set $E\subseteq(V\times{V})$.}
\KwOut {SGN$^{\textbf{(1)}}$, denoted by $G'$($V'$,$E'$).}
Initialize a node set $V'$ and a link set $E'$\;
\For {each $v \in V$}
{
    get the neighbor set $\Omega$ of $v$\;
       \For {each $\omega \in \Omega$}
          {$\ell$ = sorted([$v$, $\omega$])\;
          $\ell_{str}$ $\leftarrow$ merge the nodes in list $\ell$ into a string\;
          add the new node $\ell_{str}$ into $\widetilde{V}$\;}
       \For {$i,j \in \widetilde{V}$ and $i \neq j$}
          {add the link $(i,j)$ into $E'$\;}
    add $\widetilde{V}$ into $V'$;
}
\Return $G'$($V'$,$E'$)\;
\end{algorithm}

The process to build SGN$^{\textbf{(1)}}$ from a given network is shown in Fig.~\ref{fig:line}. In this example, the original network has 6 nodes connected by 6 links. First, extract lines as subgraphs, labeled them by their corresponding end nodes, as shown in Fig.~\ref{fig:line} (b). These lines are treated as nodes in SGN. Then, connect these lines based on their labels, i.e., two lines are connected if they share one same end node, as shown in Fig.~\ref{fig:line} (c). Finally, obtain SGN with 6 nodes and 8 links, as shown in Fig.~\ref{fig:line} (d). A pseudocode of constructing SGN$^{\textbf{(1)}}$ is given in Algorithm \ref{alg:1}. The input of this algorithm is the original network $G$($V$,$E$) and the output is the constructed SGN$^{\textbf{(1)}}$, denoted by $G'$($V'$,$E'$), where $V'$ and $E'$ represent the sets of nodes and links in the SGN$^{\textbf{(1)}}$, respectively.

\subsection{Second-Order SGN}
\begin{figure}[!t]
	\centering
	\includegraphics[width=1\linewidth]{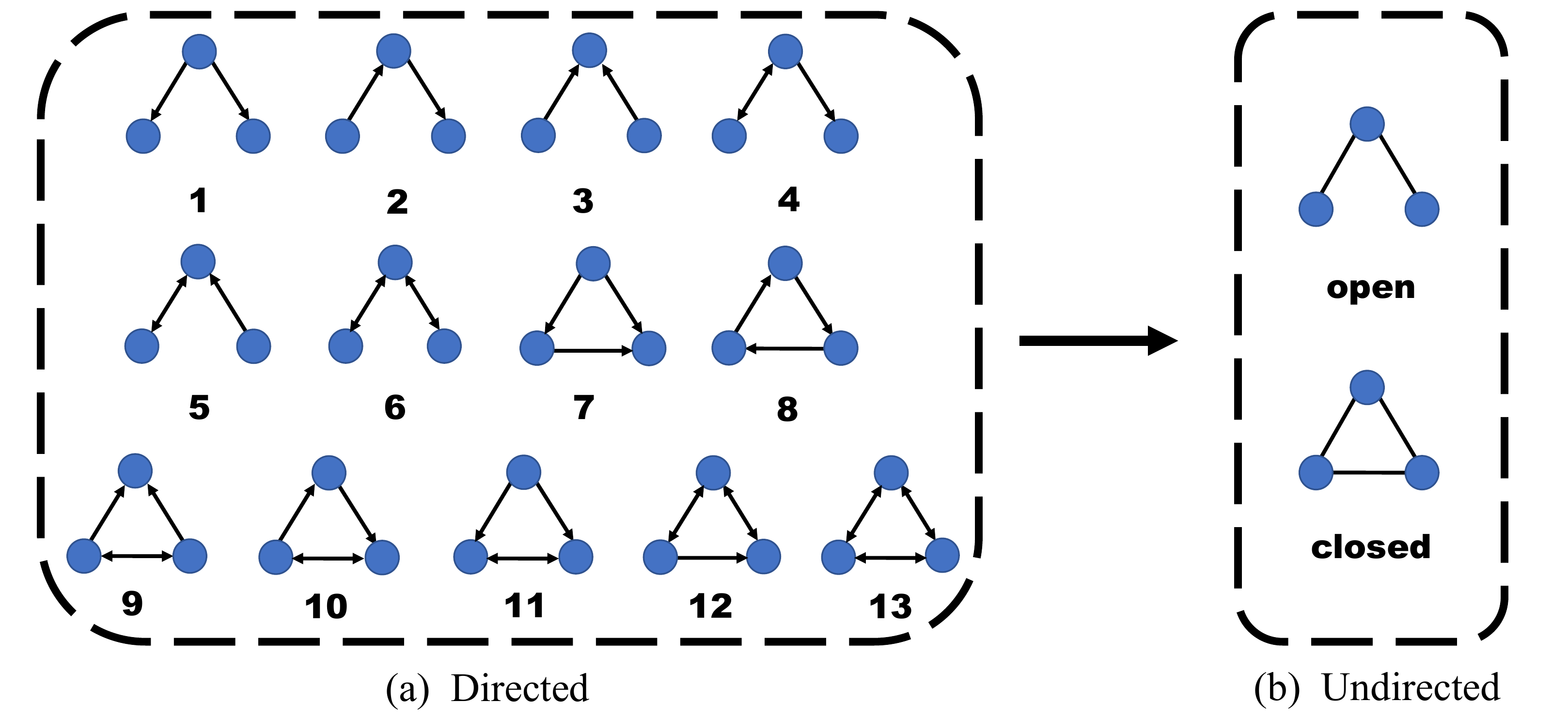}
	\caption{The connection patterns among three nodes for (a) directed and (b) undirected networks.}
	\label{fig:triangletypes}
\end{figure}

Now, construct higher-order subgraphs by considering the connection patterns among three nodes. There are more diverse connection patterns among three nodes than the case of two nodes. In theory, there are 13 possible non-isomorphic connection patterns among three nodes~\cite{wernicke2006efficient} in a directed network, as shown in Fig.~\ref{fig:triangletypes} (a). This number decreases to 2 in an undirected network, namely only open and closed triangles, as shown in Fig.~\ref{fig:triangletypes} (b). Here, only connected subgraphs are considered, while those with less than two links are ignored. Compared with lines, triangles can provide more insights about the local structure of a network~\cite{eckmann2002curvature}. For instance, in~\cite{schioberg2015evolution}, the evolution of triangles in a Google+ online social network was studied, obtaining some valuable information during the emerging and pruning of various triangles.

The open triangles are defined as subgraphs to establish the 2nd-order SGN, denoted by SGN$^{\textbf{(2)}}$. Here, second-order means that there are two links in each open triangle, and two open triangles are connected in SGN$^{\textbf{(2)}}$ if they share a same link. Note that \emph{same link} rather than \emph{same node} is used here to avoid obtaining a very dense SGN$^{\textbf{(2)}}$. This is because a dense network, with each pair of nodes connected with a higher probability, tends to provide less structural information in general.

\begin{figure}[!t]
	\centering
	\includegraphics[width=1\linewidth]{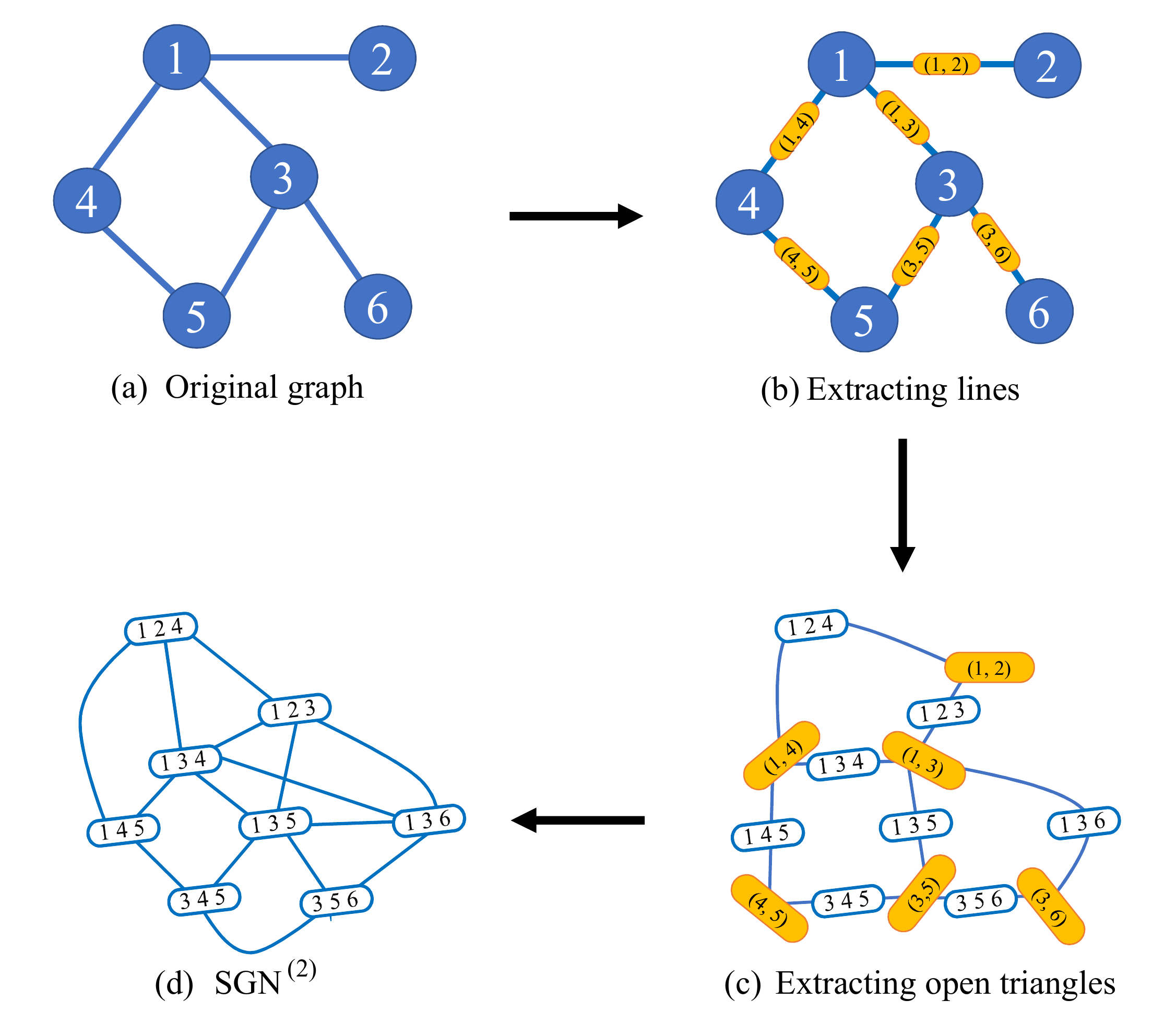}
	\caption{The process to build SGN$^{\textbf{(2)}}$ from a given network: (a) the original network, (b) extracting lines, (b) building SGN$^{\textbf{(1)}}$ and extracting open triangles as subgraphs, and (d) establishing connections among these open triangles to obtain SGN$^{\textbf{(2)}}$.}
	\label{fig:triangle}
\end{figure}

The iterative process to build SGN$^{\textbf{(2)}}$ from an original network is shown in Fig.~\ref{fig:triangle}. First, extract lines, labeled by their corresponding end nodes, as shown in Fig.~\ref{fig:triangle} (b), to establish SGN$^{\textbf{(1)}}$. Then, in the line graph SGN$^{\textbf{(1)}}$, further extract lines to obtain open triangles as subgraphs, labeled by their corresponding three nodes, as shown in Fig.~\ref{fig:triangle} (c). Finally, obtain SGN$^{\textbf{(2)}}$ with 8 nodes and 14 links, as shown in Fig.~\ref{fig:triangle} (d). A pseudocode of constructing SGN$^{\textbf{(2)}}$ is given in Algorithm \ref{alg:2}. The input of this algorithm is the original network $G$($V$,$E$) and the output is the constructed SGN$^{\textbf{(2)}}$, denoted by $G''$($V''$,$E''$), where $V''$ and $E''$ represent the sets of nodes and links in the SGN$^{\textbf{(2)}}$, respectively.

\begin{algorithm}[!t]
\caption{\textbf{Constructing SGN$^{\textbf{(2)}}$.}}
\label{alg:2}
\KwIn{A network $G$($V$,$E$) with node set $V$ and link set $E\subseteq(V\times{V})$.}
\KwOut {SGN$^{\textbf{(2)}}$, denoted by $G''$($V''$,$E''$).}
Initialize a node set $V''$ and a link set $E''$\;
\For {each $v \in V$}
{
    get the neighbors set $\Omega$ of $v$\;
    $\widetilde{\Omega}$ $\leftarrow$ get the full combination of node pairs in the neighbor collection\;
    \For {each $(\omega_{1},\omega_{2}) \in \widetilde{\Omega}$}
        {
        $\widetilde{\ell}$ = [$v,\omega_1,\omega_2$]\;
        $\widetilde{\ell}_{str}$ $\leftarrow$ merge the nodes in list $\widetilde{\ell}$ into a string\;
        add the new node $\widetilde{\ell}_{str}$ into $\widetilde{V}$\;}
         \For {$i,j \in \widetilde{V}$ and $i \neq j$}
        {add the edge ($i,j$) into $E''$ \;}
        add $\widetilde{V}$ into $V''$;
}
\Return $G''$($V''$,$E''$)\;
\end{algorithm}

Clearly, the new method can be easily extended to construct higher-order SGNs by choosing proper subgraphs and connecting rules. For instance, based on Algorithms~\ref{alg:1} and \ref{alg:2}, for the network shown in Fig.~\ref{fig:triangle} (d), one can further label each link by the 4 numbers from the end nodes, i.e., these numbers correspond to the 4 nodes in the original network. Then, one can treat each link with a different label as a node, and connect them if they share 3 same numbers, so as to establish the 3rd-order SGN.

\begin{figure}[!t]
	\centering
	\includegraphics[width=1\linewidth]{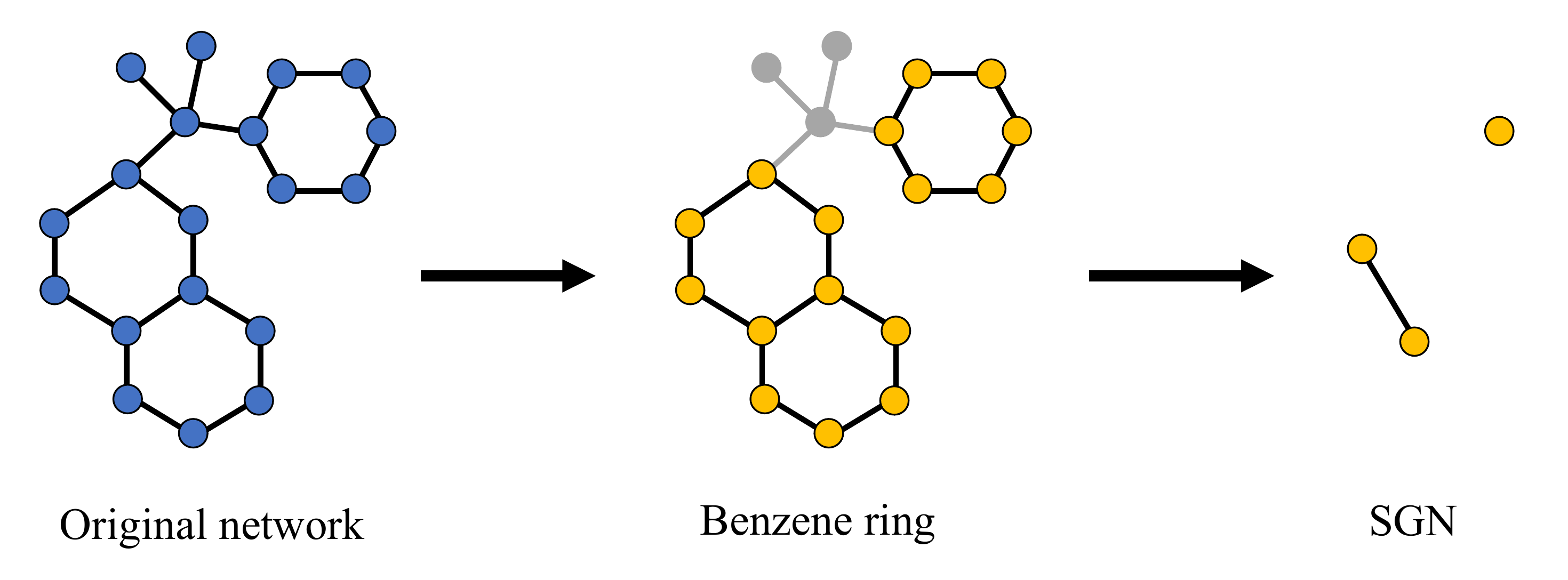}
	\caption{A compound network, where each node denotes an atom and its corresponding SGN obtained by taking benzene rings as subgraphs.}
	\label{fig:ben}
\end{figure}

It is interesting to investigate such a higher-order SGN. However, as subgraphs become too large, the SGN may contain only few nodes, making the network structure less informative. It may be argued that there might be some functional subgraphs in certain networks, which could be better blocks to be used to build SGNs. However, this may not be true. Take the compound networks in chemistry as examples, e.g. benzene ring, and other functional groups such as hydroxyl group, carboxyl group and aldehyde group, which play an important role in the properties of organic substances. In such networks, however, one usually cannot choose the benzene ring as a building block, since most of these networks are of small sizes and contain a small number of benzene rings, as shown in Fig.~\ref{fig:ben}. In this case, if one uses benzene rings as subgraphs, an SGN will be built containing only three nodes, with one isolated from the other two. As such, this SGN can hardly provide sufficient information to distinguish itself from the other substances, hence will not be useful.

\section{Network Attributes}\label{sec:NA}
Now, besides the original network, denoted by SGN$^{\textbf{(0)}}$ for simplicity, there are two SGNs, i.e., SGN$^{\textbf{(1)}}$ and SGN$^{\textbf{(2)}}$. These networks together may provide more comprehensive structural information for subsequent applications. In this paper, the focus is on its application to network classification. A typical procedure for accomplishing the task consists of two steps: first, extract network structural features; second, design a machine learning method based on these features to realize the classification. In network science, there are many classic topological attributes, which have been widely used in link prediction~\cite{wang2015link}, graph classification~\cite{li2011graph} and so on. Here, the following handcrafted network features are used to design the classifier.

\begin{itemize}
\item \textbf{Number of Nodes ($N$)}: Total number of nodes in the network.
\item \textbf{Number of links ($L$)}: Total number of links in the network.
\item \textbf{Average degree ($K$)}: The mean value of links connected to a node in the network.
\item \textbf{Percentage of leaf nodes ($P$)}: A node of degree 1 is defined as a leaf node. Suppose there are totally $F$ leaf nodes in the network. Then,
\begin{equation}
P=\frac{F}{N}\,.
\end{equation}
\item \textbf{Average clustering coefficient ($C$)}: For node $v_i$, the clustering coefficient represents the probability of a connection between any two neighbors of $v_i$. Suppose that there are $k_i$ neighbors of $v_i$ and these nodes are connected by $L_i$ links. Then, the average clustering coefficient is defined as
\begin{equation}
C=\frac{1}{N}\sum_{i=1}^N\frac{2L_i}{k_i(k_i-1)}\,.
\end{equation}
\item \textbf{Largest eigenvalue of the adjacency matrix ($\lambda$)}: The adjacency matrix $A$ of the network is an $N\times{N}$ matrix, with its element $a_{ij}=1$ if nodes $v_i$ and $v_j$ are connected, and $a_{ij}=0$ otherwise. In this step, calculate all the eigenvalues of $A$ and choose the largest one.
\item \textbf{Network density ($D$)}: Given the number of nodes $N$ and the number of links $L$, network density is defined as
\begin{equation}
D=\frac{2L}{N(N-1)}\,.
\end{equation}
\item \textbf{Average betweenness centrality ($C_B$)}: Betweenness centrality is a centrality metric based on shortest paths. The average betweenness centrality of the network is defined as
\begin{equation}
C_B=\frac{1}{N}\sum_{i=1}^N\sum_{s\neq i\neq t} {\frac{n^{i}_{st}}{g_{st}}}\,,
\end{equation}
where $g_{st}$ is the number of shortest paths between $v_s$ and $v_t$, and $n^{i}_{st}$ is the number of shortest paths between $v_s$ and $v_t$ that pass through $v_i$.  	
\item \textbf{Average closeness centrality ($C_C$)}: The closeness centrality of a node in a connected network is defined as the reciprocal of the average shortest path length between this node and the others. The average closeness centrality is defined as
\begin{equation}
C_C=\frac{1}{N}\sum_{i=1}^N\frac{{n-1}}{{\sum_{j=1}^n {{d_{ij}}} }}\,,
\end{equation}
where $d_{ij}$ is the shortest path length between nodes $v_i$ and $v_j$.
\item \textbf{Average eigenvector centrality ($C_E$)}: Usually, the importance of a node depends not only on its degree but also on the importance of its neighbors. Eigenvector centrality is another measure of the importance of a node based on its neighbors, which is defined as
\begin{equation}
C_E = \frac{1}{N} \sum_{i=1}^N x_i\,,
\end{equation}
where $x_i$ represents the importance of node $v_i$ and is calculated based on the following equation:
\begin{equation}
x_i = \alpha \sum_{j=1}^N {a_{ij}x_j}\,,
\end{equation}
where $\alpha$ is a preset parameter, which should be less than the reciprocal of the maximum eigenvalue of the adjacency matrix $A$.
\item \textbf{Average neighbor degree ($D_N$)}: Neighbor degree of a node is the average degree of all the neighbors of this node, which is defined as
\begin{equation}
D_N=\frac{1}{N}\sum_{i=1}^N \frac{1}{k_i}{\sum_{v_j\in \Omega_i} k_j}\,,
\end{equation}
where $\Omega_i$ is a set of the neighbors of node $v_i$, and $k_j$ is the degree of node $v_j\in{\Omega_i}$.
\end{itemize}

Note that, among the above 11 features, number of nodes ($N$), number of links ($L$), average degree ($K$) and network density ($D$) are the most basic properties of a network~\cite{xiaofan2012network}. Average clustering coefficient ($C$)~\cite{soffer2005network} is also a very popular metric to quantify the link density in ego networks. The percentage of leaf nodes ($P$) can distinguish whether a network is tree-like or rich with rings. The largest eigenvalue of the adjacency matrix ($\lambda$) is chosen since the eigenvalues  are the isomorphic invariant of a graph, which can be used to estimate many static attributes, such as connectivity, diameter, etc. Average neighbor degree ($D_N$) captures the 2-hop information. Also, centrality measures are indicators of the importance (status, prestige, standing, and the like) of a node in a network, therefore, we also use average betweenness centrality ($C_B$), average closeness centrality ($C_C$), and average eigenvector centrality ($C_E$) to describe the global structure of a network.

\subsection{Datasets}

Experiments were conducted on 7 real-world network datasets, as introduced in the following, with each containing two classes of networks. The first 5 datasets are about bio- and chemo-informatics, while the last two are social networks. The basic statistics of these datasets are presented in TABLE~\ref{data}.

\begin{table}[!b]\renewcommand{\arraystretch}{1.2}
	\newcommand{\tabincell}[2]{\begin{tabular}{@{}#1@{}}#2\end{tabular}}
	\caption{Basic statistics of the 7 datasets. \#Graphs is the number of graphs. \#Classes is the number of classes. \#Positive and \#Negative are the numbers of graphs in the two different classes.}
    \centering
	\begin{centering}
		\begin{tabular}{c|cccc}
			\hline\hline
			Dataset&  \#Graphs &  \#Classes&  \#Positive &  \#Negative\tabularnewline
			\hline
			MUTAG& \tabincell{c}{188}&  2&  125&  63\tabularnewline
			PTC& \tabincell{c}{344}&  2&  152&  192\tabularnewline
			PROTEINS& \tabincell{c}{1113}&  2&  663&  450\tabularnewline
			NCI1& \tabincell{c}{4110}&  2&  2057&  2053\tabularnewline
			NCI109& \tabincell{c}{4127}&  2&  2079&  2048\tabularnewline
            IMDB-B& \tabincell{c}{1000}&  2&  500&  500\tabularnewline
            REDDIT-B& \tabincell{c}{2000}&  2&  1000&  1000\tabularnewline
			\hline\hline		
		\end{tabular}		
	\end{centering}	
	\label{data}
\end{table}

\begin{itemize}
\item \textbf{\textsc{MUTAG}}: This dataset is about heteroaromatic nitro and mutagenic aromatic compounds, with nodes and links representing atoms and the chemical bonds between them, respectively. They are labeled according to whether there is a mutagenic effect on a special bacteria~\cite{debnath1991structure}.
\item \textbf{\textsc{PTC}}: This dataset includes 344 chemical compound graphs, with nodes and links representing atoms and the chemical bonds between them, respectively. Their labels are determined by their carcinogenicity for rats~\cite{toivonen2003statistical}.
\item \textbf{\textsc{PROTEINS}}: This dataset comprises of 1113 graphs. The nodes are Secondary Structure Elements (SSEs) and the links are neighbors in the amino-acid sequence or in the 3D space. These graphs represent either enzyme or non-enzyme proteins~\cite{borgwardt2005protein}.
\item \textbf{\textsc{NCI1}} \& \textbf{\textsc{NCI109}}: These two datasets comprise of 4110 and 4127 graphs, respectively. The nodes and links represent atoms and chemical bonds between them, respectively. They are two balanced subsets of the datasets of chemical compounds screened for the activities against non-small cell lung cancer and ovarian cancer cell lines, respectively. The positive and negative samples are distinguished according to whether they are effective against cancer cells~\cite{wale2008comparison}.
\item \textbf{\textsc{IMDB-B}}: This dataset is about movie collaboration, which is collected from IMDB, containing lots of information about different movies. Each graph is an ego-network, where nodes represent actors or actresses and links indicate whether they appear in the same movie. Each graph is categorized into one of the two genres (Action and Romance)~\cite{nguyen2018learning}.
\item \textbf{\textsc{REDDIT-B}}: This dataset is crawled from Reddit, which is composed of submission graphs from popular subreddits. Each graph corresponds to an online discussion thread, where nodes are users, and there is an link between two nodes if one of them responded to the other's comments. The four popular subreddits are IAmA, AskReddit, TrollXChromosomes and atheism. There are also two categories of graphs: IAmA and AskReddit are two QA-based subreddits and TrollXChromosomes and atheism are two discussion-based subreddits~\cite{yanardag2015deep}. %The task is then to identify whether a given graph belongs to a QA-based or a discussion-based community~\cite{yanardag2015deep}.}
\end{itemize}

\begin{figure}[!t]
  \centering
  \includegraphics[width=1\linewidth]{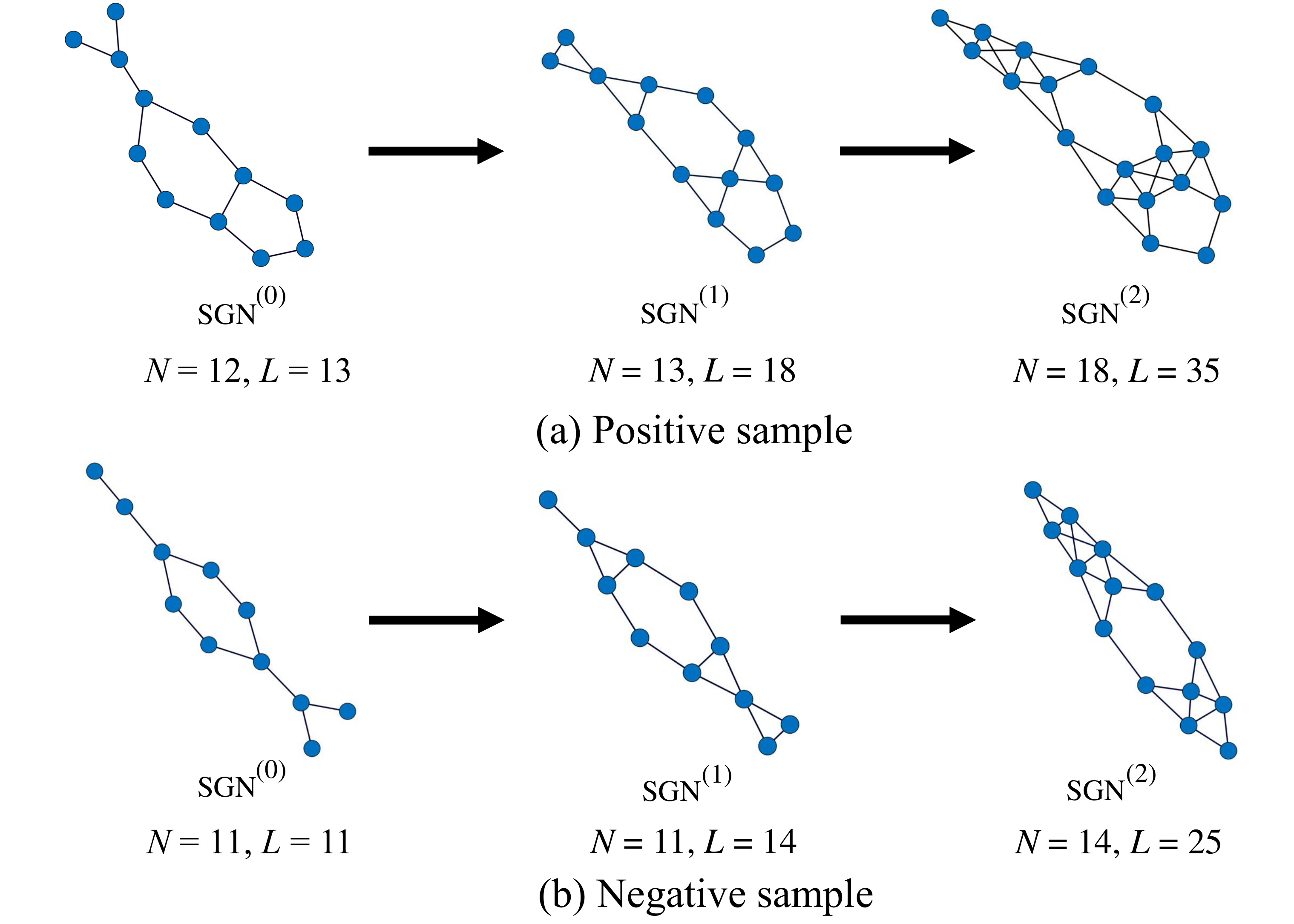}
  \caption{SGN$^{\textbf{(0)}}$, SGN$^{\textbf{(1)}}$ and SGN$^{\textbf{(2)}}$ as well as the numbers of nodes and links for (a) positive and (b) negative samples in the MUTAG dataset.}
  \label{gephi}
\end{figure}

\begin{figure*}[!t]
	\centering
	\includegraphics[width=1\linewidth]{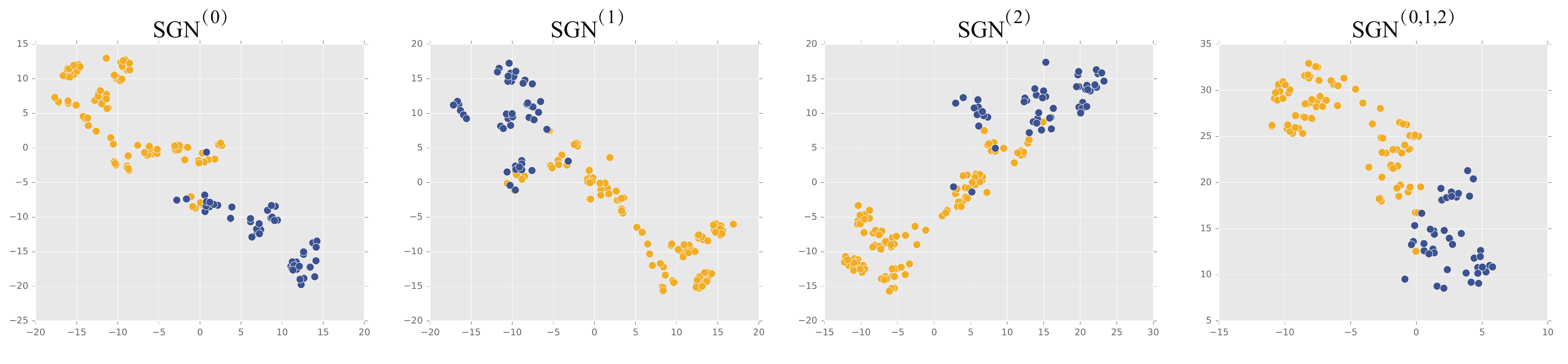}
	\caption{The t-SNE visualization of handcrafted network features. The same color of points represent the same class of networks in MUTAG.}
	\label{fig:TSNE}
\end{figure*}

\subsection{Benefits of SGN}

Here, take the MUTAG dataset as an example to show that SGNs of different orders may capture different aspects of a network structure.

First, a positive sample and a negative one are chosen from the MUTAG dataset, with their SGN$^{\textbf{(0)}}$, SGN$^{\textbf{(1)}}$ and SGN$^{\textbf{(2)}}$ visualized in Fig.~\ref{gephi}. To facilitate a comparison, the numbers of nodes and links of these networks are also presented in the figure. Here, a positive sample means that this compound has mutagenic effect on the bacteria; otherwise, it is negative. As can be seen, although the original networks of the two samples have quite similar sizes, their difference is seemingly enlarged in the higher-order SGNs; more precisely, the numbers of nodes and links in SGN increase faster for the positive sample than the negative one as the order increases.

Then, the handcrafted network features are visualized by using t-SNE in Fig.~\ref{fig:TSNE}, where the networks in MUTAG can indeed be distinguished to a certain extent by these features of the original network, the 1st-order SGN and the 2nd-order SGN, respectively. Moreover, when all the features are put together, it appears that these networks can be better distinguished, indicating that SGNs of different orders and the original network may complement to each other. Therefore, integrating the structural information of all these networks may significantly improve the performances of the subsequent algorithms designed based on network structures.

\section{Experiments}\label{sec:Exp}

With the rapid growth of real-world graph data, network classification is becoming more and more important, and a number of effective network classification methods~\cite{joachims2009predicting,kudo2005application,zhao2018substructure} have been proposed in recent years. Along this line of research, as an application of the proposed SGN, classifiers are designed based on the structural features obtained from SGNs as well as from the original networks.

\subsection{Automatic Feature Extraction Methods}

Besides those handcrafted features, one can also use some advanced methods, such as network embedding methods, to automatically generate a feature vector of certain dimension from the given network. Under the present framework, such automatically generated feature vectors can also be further expanded based on SGNs.

Network embedding method, graph2vec, and two graph kernel-based methods, subtree kernel WL and deep WL methods, and depth model algorithm CapsGNN, are chosen as automatic feature extraction methods.

\begin{itemize}
% \item \textbf{Node2vec}~\cite{grover2016node2vec}: It learns continuous feature representations in a lower-dimensional space for the nodes in a network, by optimizing the likelihood of preserving their neighborhoods. As a lower-order substructure embedding technique, it fails to learn global similarities for classifying networks directly. In the framework, first, embedding vectors are generated for all the nodes in the given network, and then the mean of them is used to represent the whole network~\cite{narayanan2017graph2vec}.

\item \textbf{Graph2vec}~\cite{narayanan2017graph2vec}: This is the first unsupervised embedding approach for an entire network, which is based on the extending word-and-document embedding techniques that has shown great advantages in NLP. Similarly, graph2vec establishes the relationship between a network and the rooted subgraphs using a similar model to \emph{doc2vec}~\cite{le2014distributed}. Graph2vec first extracts rooted subgraphs and provides corresponding labels into the vocabulary, and then trains a skip-gram model to obtain the representation of the entire network.

\item \textbf{WL}~\cite{shervashidze2011weisfeiler}: This is a rapid feature extraction scheme based on the Weisfeiler-Lehman (WL) test for isomorphism on graphs. It maps the original network to a sequence of graphs, with node attributes capturing both topological and label information. The key idea of the algorithm is to augment the node labels by the sorted set of node labels of neighboring nodes, and compress these augmented labels into new and short labels. These steps are then repeated until the node label sets of the two compared networks differ, or the number of iterations reaches a preset value. It should be noted that, to facilitate the expansion of the new model, the sub-structure frequency vectors, instead of the kernel matrix $\mathcal{K}$, are used as the inputs to the new classifier.

\item \textbf{Deep WL}~\cite{yanardag2015deep}: This provides a unified framework that leverages the dependency information of sub-structures by learning latent representations. The differences from the WL kernel generate a corpus of sub-structures by integrating language-modeling and deep-learning techniques~\cite{bengio2003neural}, where a co-occurrence relationship of sub-structures is preserved and sub-structure vector representations are obtained before the kernel is computed. Then, a sub-structure similarity matrix, $\mathcal{M}$, is calculated by the matrix $\mathcal{V}$ with each column representing a sub-structure vector. Denote by $\mathcal{P}$ the matrix with each column representing a sub-structure frequency vector. Then, according to the definition of kernel:
\begin{equation}
\mathcal{K} = \mathcal{P}\mathcal{M}\mathcal{P}^\mathrm{T} = \mathcal{P}\mathcal{V}\mathcal{V}^\mathrm{T}\mathcal{P}^\mathrm{T}=\mathcal{H}\mathcal{H}^\mathrm{T},
\end{equation}
one can use the columns in the matrix $\mathcal{H}=\mathcal{P}\mathcal{V}$ as the inputs to the classifier.

\item \textbf{CapsGNN}~\cite{xinyi2018capsule}: This method was inspired by CapsNet, which adopted the concept of capsules to overcome the weakness of existing GNN-based graph embedding algorithms. In particular, CapsGNN extracts node features in the form of capsules and utilizes the routing mechanism to capture important information at the graph level. The model generates multiple embeddings for each graph so as to capture graph properties from different aspects.
\end{itemize}

% In this study, the same embedding dimension is adopted for graph2vec and node2vec according to ~\cite{narayanan2017graph2vec}.
% For \emph{node2vec}, the following default parameters are used: the number of random walks is set to 10, the walk length is set to 80, the window size is set to 10 for the skip-gram model, and the embedding dimension is set to 1024. In this method, the biased random walk is determined by two hyper-parameters, $p$ and $q$, both are set to 1 for simplicity.

In this study, for \emph{graph2vec}, the embedding dimension is adopted  according to~\cite{narayanan2017graph2vec}. Graph2vec is based on the rooted subgraphs which are adopted in the WL kernel. The parameter height of \emph{WL kernel} is set to 3. Since the embedding dimension is predominant for learning performances, a commonly-used value of 1024 is adopted. The other parameters are set to defaults: the learning rate is set to 0.5, the batch size is set to 512 and the epochs is set to 1000. For \emph{WL} and \emph{Deep WL}, according to ~\cite{yanardag2015deep}, the Weisfelier-Lehman subtree kernel is used to built the corpus and the height of which is set to 2. Then, the Maximum Likelihood Estimation (MLE) is used to compute the kernel in the WL method. Furthermore, the same parameter setting as WL is chosen, with the embedding dimension equal to 10, window size equal to 5 and skip-gram used for the word2vec model in the deep WL method. We adopt the default parameters for \emph{CapsGNN} and flatten the multiple embeddings of each graph as the input.

Without loss of generality, the well-known logistic regression is chosen as the new classification model. Meanwhile, for each feature extraction method, the feature space is first expanded by using SGNs, and then the dimension of the feature vectors is reduced to the same value as that of the feature vector obtained from the original network using PCA in the experiments, for a fair comparison. Each dataset is randomly split into 9 folds for training and 1 fold for testing. Here, the $F_1$-$Score$ is adopted as the metric to evaluate the classification performance:
\begin{equation}
F_{1}= \frac{2PR}{P + R}\,,
\end{equation}
where $P$ and $R$ are the precision and recall, respectively. To exclude the random effect of the fold assignment, experiment is repeated for 500 times and then the average $F_1$-$Score$ and its standard deviation are recorded.

\begin{table*}[!t]
\caption{Classification results on the 7 datasets, in terms of $F_1$-$Score$, based on different feature extraction methods and combinations of SGNs.}
\centering
\renewcommand\arraystretch{1.1}
\begin{tabular}{c|cccccccc}
\hline\hline
\multicolumn{2}{c|}{ $\textbf{Algorithm}$}& \multicolumn{7}{c}{Dataset} \\
\hline
\multicolumn{2}{c|}{Handcraft}&MUTAG&PTC&PROTEINS&NCI1&NCI109&IMDB-B&REDDIT-B\\
\hline
\multicolumn{2}{c|}{SGN$^{\textbf{(0)}}$}&$86.58\pm{3.61}$&$63.52\pm{4.55}$&$78.30\pm{2.49}$&$67.48\pm{0.87}$&$67.34\pm{1.25}$&$73.00\pm{3.68}$&
$78.68\pm{1.66}$\\
\multicolumn{2}{c|}{SGN$^{\textbf{(1)}}$}&$88.20\pm{3.62}$&$65.29\pm{6.93}$&$76.79\pm{3.41}$&$65.72\pm{1.41}$&$66.25\pm{2.14}$&$73.30\pm{4.82}$&
$76.50\pm{2.73}$\\
\multicolumn{2}{c|}{SGN$^{\textbf{(2)}}$}&$85.53\pm{4.47}$&$65.00\pm{6.09}$&$75.45\pm{5.04}$&$65.35\pm{2.32}$&$64.15\pm{2.20}$&$74.24\pm{3.38}$&
$74.37\pm{3.14}$\\
\multicolumn{2}{c|}{SGN$^{\textbf{(01)}}$}&$87.89\pm{4.58}$&$66.47\pm{6.73}$&$78.83\pm{3.12}$&$68.76\pm{2.24}$&$68.88\pm{2.17}$&$73.38\pm{3.94}$&
$79.15\pm{2.32}$\\
\multicolumn{2}{c|}{SGN$^{\textbf{(0,2)}}$}&$88.42\pm{4.22}$&$65.59\pm{7.09}$&$78.92\pm{3.17}$&$69.39\pm{1.82}$&$68.09\pm{1.74}$&$75.42\pm{3.34}$&
$78.80\pm{2.07}$\\
\multicolumn{2}{c|}{SGN$^{\textbf{(1,2)}}$}&$88.95\pm{3.37}$&$67.06\pm{6.14}$&$78.21\pm{3.60}$&$68.13\pm{1.30}$&$68.48\pm{1.42}$&$74.94\pm{2.81}$&
$77.23\pm{1.98}$\\
\multicolumn{2}{c|}{SGN$^{\textbf{(0,1,2)}}$}&$\textbf{91.58}\pm{4.21}$&$\textbf{67.94}\pm{6.36}$&  $\textbf{79.46}\pm{2.96}$&  $\textbf{69.84}\pm{1.59}$&  $\textbf{69.73}\pm{1.97}$&  $\textbf{77.65}\pm{4.50}$&
$\textbf{79.23}\pm{1.62}$\\
% \multicolumn{2}{c|}{SGN$^{\textbf{(0,1,2,3)}}$}&$\textbf{91.58}\pm{6.32}$&$67.64\pm{6.22}$&  $79.21\pm{3.42}$&  $69.50\pm{1.32}$&  $\textbf{71.20}\pm{1.19}$&  $76.72\pm{3.89}$&
% $79.20\pm{2.96}$\\
\hline
% \multicolumn{2}{c|}{Gain}&5.78\%&  6.96\%&  1.71\%&  3.50\%&  5.73\%&  6.37\%&  0.70\%\\
\multicolumn{2}{c|}{Gain}&5.78\%&  6.96\%&  1.71\%&  3.50\%&  3.55\%&  6.37\%&  0.70\%\\
\hline
% \multicolumn{2}{c|}{Node2vec}&MUTAG&PTC&PROTEINS&NCI1&NCI109&IMDB-BINARY\\
% \hline
% \multicolumn{2}{c|}{SGN$^{\textbf{(0)}}$}&$72.63\pm{10.20}$&$58.85\pm{8.00}$&$57.49\pm{3.57}$&$54.89\pm{1.61}$&$52.68\pm{1.56}$&$55.53\pm{2.15}$\\
% \multicolumn{2}{c|}{SGN$^{\textbf{(1)}}$}&$82.89\pm{5.67}$&$55.59\pm{4.25}$&$67.44\pm{2.17}$&$\textbf{62.31}\pm{0.92}$&$61.31\pm{1.50}$&$56.82\pm{2.12}$\\
% \multicolumn{2}{c|}{SGN$^{\textbf{(2)}}$}&$85.53\pm{3.38}$&$54.85\pm{3.84}$&$65.39\pm{4.29}$&$56.74\pm{0.92}$&$57.55\pm{1.49}$&$56.47\pm{3.48}$\\
% \multicolumn{2}{c|}{SGN$^{\textbf{(0,1)}}$}&$83.42\pm{3.73}$&$\textbf{60.74}\pm{4.16}$&$67.43\pm{3.50}$&$62.14\pm{1.29}$&$\textbf{62.40}\pm{1.08}$&$58.00\pm{1.45}$\\
% \multicolumn{2}{c|}{SGN$^{\textbf{(0,2)}}$}&$87.11\pm{6.17}$&$55.15\pm{7.09}$&$67.62\pm{2.39}$&$61.20\pm{1.25}$&$61.89\pm{1.67}$&$58.17\pm{1.40}$\\
% \multicolumn{2}{c|}{SGN$^{\textbf{(1,2)}}$}&$86.47\pm{4.17}$&$55.75\pm{3.85}$&$67.76\pm{3.46}$&$61.70\pm{1.33}$&$60.82\pm{1.42}$&$56.94\pm{2.95}$\\
% \multicolumn{2}{c|}{SGN$^{\textbf{(0,1,2)}}$}&$\textbf{87.39}\pm{5.50}$&$58.38\pm{5.66}$&$\textbf{67.94}\pm{2.62}$&$61.48\pm{0.88}$&  $61.30\pm{1.58}$&$\textbf{58.42}\pm{1.47}$\\
% \hline
% \multicolumn{2}{c|}{Gain}&20.32\%&  3.69\%&  18.18\%&  13.52\%&  18.45\%&  5.20\%\\
\hline
\multicolumn{2}{c|}{Graph2vec}&MUTAG&PTC&PROTEINS&NCI1&NCI109&IMDB-B&REDDIT-B\\
\hline
\multicolumn{2}{c|}{SGN$^{\textbf{(0)}}$}&$83.15\pm{9.25}$&$60.17\pm{6.86}$&$73.30\pm{2.05}$&$73.22\pm{1.81}$&  $74.26\pm{1.47}$&$62.47\pm{3.99}$&
$76.00\pm{2.20}$\\
\multicolumn{2}{c|}{SGN$^{\textbf{(1)}}$}&$63.16\pm{4.68}$&$56.80\pm{5.39}$&$60.27\pm{2.05}$&$54.56\pm{1.38}$&  $56.35\pm{1.52}$&$63.06\pm{6.72}$&
$75.34\pm{2.55}$\\
\multicolumn{2}{c|}{SGN$^{\textbf{(2)}}$}&$68.95\pm{8.47}$&$57.35\pm{3.83}$&$59.82\pm{4.11}$&$61.31\pm{2.13}$&  $53.54\pm{1.43}$&$64.35\pm{6.63}$&
$74.50\pm{2.71}$\\
\multicolumn{2}{c|}{SGN$^{\textbf{(0,1)}}$}&$83.42\pm{5.40}$&$59.03\pm{3.36}$&$74.12\pm{1.57}$&$73.65\pm{1.38}$&$73.18\pm{1.26}$&$66.59\pm{4.54}$&
$77.63\pm{1.25}$\\
\multicolumn{2}{c|}{SGN$^{\textbf{(0,2)}}$}&$81.32\pm{3.80}$&$61.76\pm{3.73}$&$73.09\pm{1.28}$&  $\textbf{77.54}\pm{2.52}$&$\textbf{75.39}\pm{1.33}$&$66.53\pm{4.45}$&
$77.39\pm{3.10}$\\
\multicolumn{2}{c|}{SGN$^{\textbf{(1,2)}}$}&$72.63\pm{4.08}$&$59.42\pm{5.84}$&$62.76\pm{3.49}$&$67.47\pm{3.84}$&$68.12\pm{1.86}$&$66.24\pm{2.58}$&
$76.00\pm{3.18}$\\
\multicolumn{2}{c|}{SGN$^{\textbf{(0,1,2)}}$}&$\textbf{86.84}\pm{5.70}$&$\textbf{63.24}\pm{6.70}$&  $\textbf{74.44}\pm{3.09}$&$76.64\pm{3.21}$&$74.86\pm{2.76}$&$\textbf{70.65}\pm{5.55}$&
$\textbf{78.04}\pm{2.61}$\\
% \multicolumn{2}{c|}{SGN$^{\textbf{(0,1,2,3)}}$}&$\textbf{87.72}\pm{4.04}$&$\textbf{64.53}\pm{3.97}$&  $\textbf{74.88}\pm{3.20}$&$\textbf{78.34}\pm{2.68}$&$\textbf{76.85}\pm{0.44}$&$69.49\pm{4.86}$&
% $\textbf{78.10}\pm{3.04}$\\
\hline
% \multicolumn{2}{c|}{Gain}&5.50\%&  7.25\%&  2.16\%&  7.00\%&  3.49\%&  13.73\%&  2.76\%\\
\multicolumn{2}{c|}{Gain}&4.44\%&  5.10\%&  1.56\%&  5.90\%&  1.52\%&  13.73\%&  2.68\%\\
\hline
\multicolumn{2}{c|}{WL}&MUTAG&PTC&PROTEINS&NCI1&NCI109&IMDB-B&REDDIT-B\\
\hline
\multicolumn{2}{c|}{SGN$^{\textbf{(0)}}$}&$80.63\pm{3.07}$&  $56.91\pm{2.79}$&  $72.92\pm{0.56}$&  $66.19\pm{0.97}$&  $69.26\pm{1.14}$&$70.90\pm{4.18}$&
$75.15\pm{2.39}$\\
\multicolumn{2}{c|}{SGN$^{\textbf{(1)}}$}&$76.05\pm{2.73}$&  $61.76\pm{6.17}$&  $67.04\pm{1.58}$&  $58.24\pm{1.87}$&  $57.68\pm{1.28}$&$69.20\pm{3.87}$&
$74.83\pm{2.64}$\\
\multicolumn{2}{c|}{SGN$^{\textbf{(2)}}$}&$74.21\pm{7.33}$&  $59.41\pm{5.22}$&  $64.01\pm{1.58}$&  $52.62\pm{0.53}$&  $58.26\pm{0.72}$&$66.10\pm{4.18}$&
$74.34\pm{2.65}$\\
\multicolumn{2}{c|}{SGN$^{\textbf{(0,1)}}$}&$87.11\pm{5.45}$&  $62.50\pm{4.15}$&  $76.19\pm{2.29}$&  $72.49\pm{1.79}$&  $69.50\pm{1.76}$&$73.05\pm{4.75}$&
$76.90\pm{1.51}$\\
\multicolumn{2}{c|}{SGN$^{\textbf{(0,2)}}$}&$86.57\pm{4.31}$&  $59.71\pm{3.96}$&  $74.99\pm{1.56}$&  $70.11\pm{1.22}$&  $69.67\pm{1.34}$&$72.23\pm{3.13}$&
$75.40\pm{4.73}$\\
\multicolumn{2}{c|}{SGN$^{\textbf{(1,2)}}$}&$76.11\pm{7.75}$&  $60.88\pm{3.11}$&  $64.81\pm{3.05}$&  $56.00\pm{2.35}$&  $57.92\pm{1.30}$&$70.45\pm{3.22}$&
$74.15\pm{4.17}$\\
\multicolumn{2}{c|}{SGN$^{\textbf{(0,1,2)}}$}&$\textbf{88.94}\pm{3.28}$&  $\textbf{63.53}\pm{4.80}$&  $\textbf{78.08}\pm{1.41}$&  $\textbf{77.03}\pm{2.73}$&  $\textbf{72.92}\pm{1.25}$&$\textbf{75.00}\pm{4.49}$&
$\textbf{77.03}\pm{2.73}$\\
% \multicolumn{2}{c|}{SGN$^{\textbf{(0,1,2,3)}}$}&$\textbf{89.47}\pm{6.57}$&  $\textbf{64.63}\pm{5.26}$&  $77.32\pm{3.20}$&  $73.11\pm{1.56}$&  $\textbf{72.95}\pm{1.42}$&$73.58\pm{5.26}$&
% $\textbf{78.10}\pm{2.70}$\\
\hline
% \multicolumn{2}{c|}{Gain}&10.96\%&  13.57\%&  7.08\%&  11.63\%&  5.33\%&  5.78\%&  3.93\%\\
\multicolumn{2}{c|}{Gain}&10.31\%&  11.63\%&  7.08\%&  11.63\%&  5.28\%&  5.78\%&  2.50\%\\
\hline
\multicolumn{2}{c|}{Deep WL}&MUTAG&PTC&PROTEINS&NCI1&NCI109&IMDB-B&REDDIT-B\\
\hline
\multicolumn{2}{c|}{SGN$^{\textbf{(0)}}$}&$82.95\pm{2.68}$&  $59.04\pm{1.09}$&  $73.30\pm{0.82}$&  $67.06\pm{1.91}$&  $67.04\pm{1.36}$&  $67.50\pm{2.45}$&
$77.25\pm{2.52}$\\
\multicolumn{2}{c|}{SGN$^{\textbf{(1)}}$}&$67.89\pm{6.84}$&  $58.53\pm{3.23}$&  $69.43\pm{2.57}$&  $55.45\pm{1.43}$&  $57.63\pm{2.07}$&  $73.30\pm{2.38}$&
$76.93\pm{3.68}$\\
\multicolumn{2}{c|}{SGN$^{\textbf{(2)}}$}&$68.42\pm{6.65}$&  $62.65\pm{4.17}$&  $68.57\pm{2.42}$&  $55.22\pm{1.45}$&  $55.68\pm{1.12}$&  $71.48\pm{2.48}$&
$75.29\pm{4.35}$\\
\multicolumn{2}{c|}{SGN$^{\textbf{(0,1)}}$}&$92.11\pm{5.39}$&  $64.41\pm{1.87}$&  $74.62\pm{2.51}$&  $70.10\pm{1.24}$&  $69.39\pm{1.35}$&  $73.50\pm{2.87}$&
$77.35\pm{2.47}$\\
\multicolumn{2}{c|}{SGN$^{\textbf{(0,2)}}$}&$93.15\pm{5.28}$&  $64.70\pm{4.88}$&  $75.89\pm{2.99}$&  $70.12\pm{1.31}$&  $68.61\pm{1.11}$&  $74.36\pm{2.22}$&
$76.92\pm{3.13}$\\
\multicolumn{2}{c|}{SGN$^{\textbf{(1,2)}}$}&$73.68\pm{5.77}$&  $64.16\pm{4.92}$&  $69.43\pm{2.26}$&  $59.95\pm{1.12}$&  $56.24\pm{1.62}$&  $71.90\pm{2.51}$&
$76.20\pm{5.06}$\\
\multicolumn{2}{c|}{SGN$^{\textbf{(0,1,2)}}$}&$\textbf{93.68}\pm{5.15}$&  $\textbf{65.88}\pm{5.05}$&  $\textbf{76.78}\pm{2.41}$&  $\textbf{70.26}\pm{1.24}$&  $\textbf{71.06}\pm{1.61}$&  $\textbf{75.70}\pm{1.55}$&
$\textbf{78.41}\pm{1.70}$\\
% \multicolumn{2}{c|}{SGN$^{\textbf{(0,1,2,3)}}$}&$\textbf{93.74}\pm{2.46}$&  $\textbf{68.35}\pm{2.96}$&  $75.97\pm{1.44}$&  $\textbf{71.52}\pm{1.40}$&  $70.57\pm{1.26}$&  $\textbf{76.00}\pm{2.53}$&
% $\textbf{79.00}\pm{1.95}$\\
\hline
% \multicolumn{2}{c|}{Gain}&13.01\%&  15.77\%&  4.75\%&  6.65\%&  6.00\%&  12.59\%&  2.27\%\\
\multicolumn{2}{c|}{Gain}&12.93\%&  11.58\%&  4.75\%&  4.77\%&  6.00\%&  11.85\%&  1.50\%\\
\hline
\multicolumn{2}{c|}{CapsGNN}&MUTAG&PTC&PROTEINS&NCI1&NCI109&IMDB-B&REDDIT-B\\
\hline
\multicolumn{2}{c|}{SGN$^{\textbf{(0)}}$}&$86.32\pm{7.52}$&  $62.06\pm{4.25}$&  $75.89\pm{3.51}$&  $78.30\pm{1.80}$&  $72.99\pm{2.15}$&  $72.71\pm{4.36}$&
$76.12\pm{3.82}$\\
\multicolumn{2}{c|}{SGN$^{\textbf{(1)}}$}&$83.68\pm{8.95}$&  $61.76\pm{5.00}$&  $74.64\pm{3.55}$&  $74.70\pm{1.54}$&  $69.82\pm{2.24}$&  $74.35\pm{4.61}$&
$75.64\pm{4.15}$\\
\multicolumn{2}{c|}{SGN$^{\textbf{(2)}}$}&$82.63\pm{7.08}$&  $58.82\pm{3.95}$&  $72.39\pm{6.03}$& $69.82\pm{1.89}$&  $67.04\pm{2.45}$&  $73.64\pm{4.77}$&
$72.41\pm{3.97}$\\
\multicolumn{2}{c|}{SGN$^{\textbf{(0,1)}}$}&$87.37\pm{8.55}$&  $63.53\pm{4.40}$&  $76.25\pm{3.53}$&  $78.42\pm{2.92}$&  $73.28\pm{3.11}$&  $74.58\pm{4.80}$&
$78.49\pm{2.93}$\\
\multicolumn{2}{c|}{SGN$^{\textbf{(0,2)}}$}&$87.89\pm{5.29}$&  $62.20\pm{6.14}$&  $73.00\pm{3.17}$& $73.78\pm{2.32}$&  $71.52\pm{2.09}$&  $74.94\pm{4.56}$&
$78.17\pm{5.13}$\\
\multicolumn{2}{c|}{SGN$^{\textbf{(1,2)}}$}&$78.95\pm{8.49}$&  $59.11\pm{5.65}$&  $70.09\pm{2.45}$&  $70.53\pm{2.45}$&  $70.64\pm{2.30}$&  $75.29\pm{4.08}$&
$75.73\pm{4.97}$\\
\multicolumn{2}{c|}{SGN$^{\textbf{(0,1,2)}}$}&$\textbf{89.47}\pm{7.44}$& $\textbf{64.12}\pm{3.67}$&  $\textbf{76.34}\pm{4.13}$&  $\textbf{78.61}\pm{1.87}$& $\textbf{73.72}\pm{2.39}$&  $\textbf{76.47}\pm{5.74}$&
$\textbf{79.68}\pm{5.34}$\\
% \multicolumn{2}{c|}{SGN$^{\textbf{(0,1,2,3)}}$}&$\textbf{93.74}\pm{2.46}$&  $\textbf{68.35}\pm{2.96}$&  $75.97\pm{1.44}$&  $\textbf{71.52}\pm{1.40}$&  $70.57\pm{1.26}$&  $\textbf{76.00}\pm{2.53}$&
% $\textbf{79.00}\pm{1.95}$\\
\hline
% \multicolumn{2}{c|}{Gain}&13.01\%&  15.77\%&  4.75\%&  6.65\%&  6.00\%&  12.59\%&  2.27\%\\
\multicolumn{2}{c|}{Gain}&4.65\%& 3.32\%&  0.59\%&  0.40\%&  1.00\%&  5.17\%&  4.87\%\\

\hline\hline
\end{tabular}
\label{Table2}
\end{table*}

\subsection{Computational Complexity}
Now, the computational complexity in building SGNs is analyzed. Denote by $|V|$ and $|E|$ the numbers of nodes and links, respectively, in the original network. The average degree of the network is calculated by
\begin{equation}
K=\frac{1}{|V|}\sum_{i=1}^{|V|}k_i=\frac{2|E|}{|V|}\,,
\end{equation}
where $k_i$ is the degree of node $v_i$. Based on Algorithm \ref{alg:1}, the time complexity in transforming the original network to SGN$^{\textbf{(1)}}$ is
\begin{equation}
\mathcal{T}_1=\mathcal{O}(K|V|+|E|^2)=\mathcal{O}(|E|^2+|E|) = \mathcal{O}(|E|^2)\,.
\end{equation}
Then, the number of nodes in SGN$^{\textbf{(1)}}$ is equal to $|E|$ and the number of links is $\sum_{i=1}^{|V|}k_i^2-|E|\leq|E|^2-|E|$ \cite{harary1960some}. Similarly, one can get the time complexity in transforming SGN$^{\textbf{(1)}}$ to SGN$^{\textbf{(2)}}$, as
\begin{equation}
\mathcal{T}_2 \leq \mathcal{O}((|E|^2-|E|)^2) = \mathcal{O}(|E|^4)\,.
\end{equation}

\begin{figure*}[!t]
  \centering
  \includegraphics[width=1\linewidth]{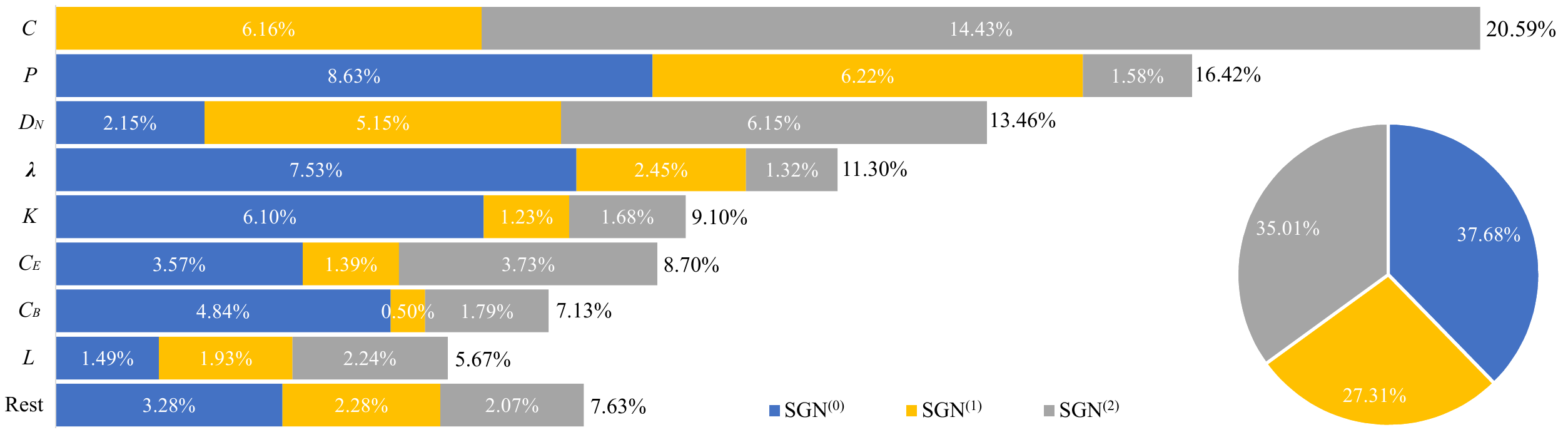}
  \caption{The importance of handcrafted features in logistic regression model for network classification using SGN$^{\textbf{(0)}}$, SGN$^{\textbf{(1)}}$ and SGN$^{\textbf{(2)}}$ together in MUTAG dataset.}
  \label{featureout}
\end{figure*}

\begin{figure*}[!t]
  \centering
  \includegraphics[width=1\linewidth]{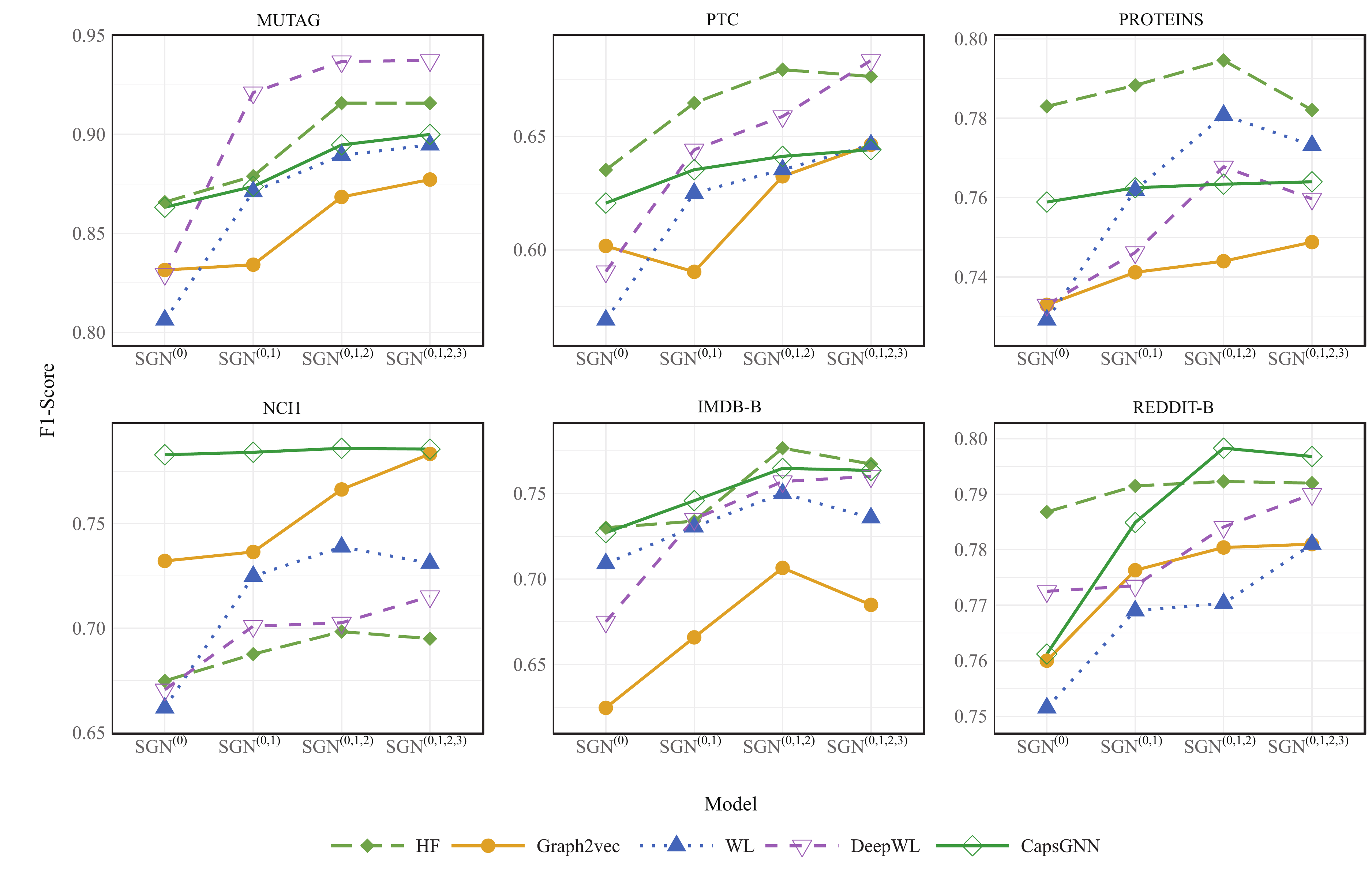}
  \caption{Average F1-Scores obtained by using different feature extraction methods under different combinations of SGNs.}
  \label{fig:exps}
\end{figure*}

\subsection{Experiment Results}
As described in Sec.~\ref{sec:SGN}, the proposed SGNs can be used to expand structural feature spaces. To investigate the effectiveness of the 1st-order and the 2nd-order SGNs, i.e., SGN$^{\textbf{(1)}}$ and SGN$^{\textbf{(2)}}$, for each feature extraction method, the classification results are compared on the basis of only one network, i.e., SGN$^{\textbf{(0)}}$, SGN$^{\textbf{(1)}}$ and SGN$^{\textbf{(2)}}$, respectively; on the basis of two networks, i.e., SGN$^{\textbf{(0)}}$ together with SGN$^{\textbf{(1)}}$ and SGN$^{\textbf{(0)}}$ together with SGN$^{\textbf{(2)}}$, denoted by SGN$^{\textbf{(0,1)}}$ and SGN$^{\textbf{(0,2)}}$, respectively; and on the basis of three networks, i.e., SGN$^{\textbf{(0)}}$ together with SGN$^{\textbf{(1)}}$ and SGN$^{\textbf{(2)}}$, denoted as SGN$^{\textbf{(0,1,2)}}$. For a fair comparison, PCA is used to compress the feature vectors to the same dimension for each feature extraction method, before they are input into the logistic regression model.

The results are shown in TABLE~\ref{Table2}, where one can see that, for a single network case, the original network seems to provide more structural information, i.e., the classification model based on SGN$^{\textbf{(0)}}$ performs better, in terms of higher $F_1$-$Score$, than those based on SGN$^{\textbf{(1)}}$ or SGN$^{\textbf{(2)}}$, in most cases. This is reasonable, because there must be information loss in the processes to build SGNs. However, it still appears to be dependent on the feature extraction method used. For instance, when the Deep WL is adopted, better classification results can be obtained based on SGN$^{\textbf{(1)}}$ or SGN$^{\textbf{(2)}}$ than SGN$^{\textbf{(0)}}$ for 2 datasets, while when handcrafted features are used, even better classification performance is realized based on the 1st-order or 2nd-order SGNs than the original network in 3 datasets. More interestingly, the classification models based on two networks, i.e., SGN$^{\textbf{(0,1)}}$ and SGN$^{\textbf{(0,2)}}$, perform better than those based on a single network, while the model based on three networks, i.e., SGN$^{\textbf{(0,1,2)}}$, performs the best in most cases.

The gain $\mathcal{G}$ on $F_1$-$Score$ is calculated, when all the three networks are used together, i.e., SGN$^{\textbf{(0,1,2)}}$, compared with that when only the original network is used, i.e., SGN$^{\textbf{(0)}}$, which is defined to be the relatively difference between their corresponding $F_1$-$Score$:
\begin{equation}
\mathcal{G}=\frac{F_1^{(0,1,2)}-F_1^{(0)}}{F_1^{(0)}}\times{100\%}\,.
\end{equation}
The gains are also presented in TABLE~\ref{Table2}, where one can see that the classification performance is indeed significantly improved in all the 35 cases. Particularly, in 17 cases, the gains are larger than 5\%, while in 7 cases, they are even larger than 10\%. These results indicate that the 1st-order and the 2nd-order SGNs can indeed complement the original network regarding the structural information, thus benefiting network classification. Surprisingly, it is found that the chosen handcrafted features based on SGN$^{\textbf{(0,1,2)}}$ outperforms the other automatically generated features that use more advanced network-embedding or graph-kernel based methods even depth model, in 3 out of 7 datasets, i.e., PTC, PROTEINS and IMDB-B. This phenomenon indicates that, compared with those automatically generated ones, properly chosen traditional structural features are of particular advantage in the proposed framework, in the sense that they are not only more interpretable due to their clear physical meanings, but also equally effective in designing subsequent structure-based algorithms, e.g., for network classification.

In addition, the feature importance for the task of network classification is investigated by using logistic regression. Denote by $\beta_i$ the coefficient of feature $x_i$ in the model, and suppose that there are $M$ features in total. Then, the importance of feature $x_i$ is defined as
\begin{equation}
\mathcal{I}=\frac{|\beta_i|}{\sum_{k=1}^M{|\beta_k|}}\times{100\%}\,.
\end{equation}

Taking MUTAG for example, the results are visualized in Fig.~\ref{featureout}. Overall, the features in SGN$^{\textbf{(0)}}$ are most important, since they determine 37.68\% of the model, while the features in SGN$^{\textbf{(2)}}$ are more important than those in SGN$^{\textbf{(1)}}$, since they determine 35.01\% and 27.31\% of the model, respectively. When focusing on a single feature, it is found that the clustering coefficient $C$, the percentage of leaf nodes $P$, and the average neighbor degree $D_N$, are the top three most important features, and they together determine more than 50\% of the model. Interestingly, it appears that different SGNs address different aspects of the network structure in the classification task. For instance, the most important feature in SGN$^{\textbf{(2)}}$ is the clustering coefficient, while the coefficient for this feature in SGN$^{\textbf{(0)}}$ is zero since there is no triangle in the networks in MUTAG dataset. Moreover, the largest eigenvalue of the adjacency matrix $\lambda$ and the average degree $K$ in SGN$^{\textbf{(0)}}$ are relatively important, while those in SGN$^{\textbf{(1)}}$ and SGN$^{\textbf{(2)}}$ have less effect on the model. These results confirm once again that SGNs indeed complement the original network to achieve better network classification performance.

\begin{figure}[!t]
  \centering
  \includegraphics[width=1\linewidth]{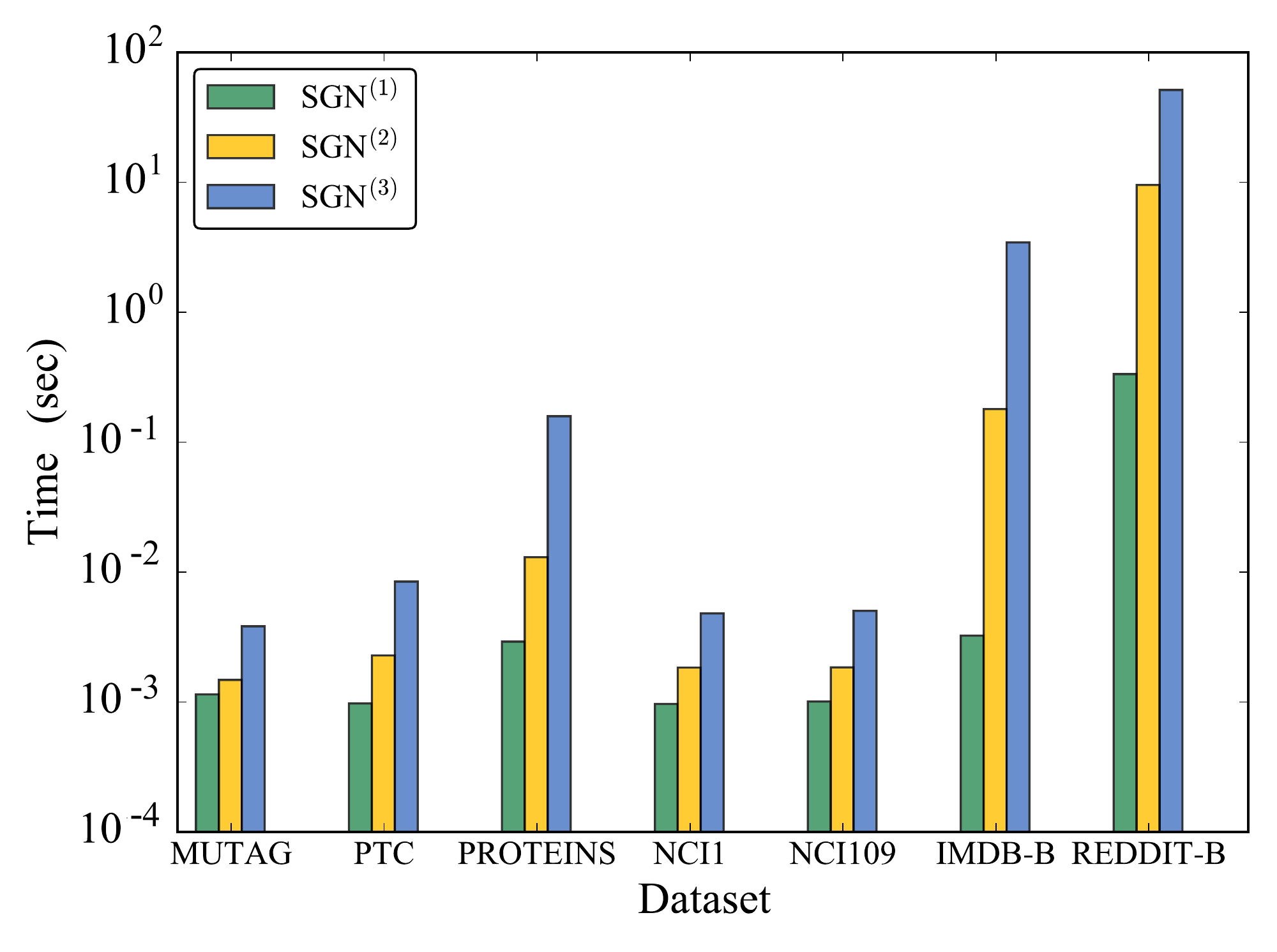}
  \caption{Average execution time to establish SGNs of different orders on the seven datasets.}
  \label{fig:time}
\end{figure}

\begin{figure*}[!t]
	\centering
	\includegraphics[width=\linewidth]{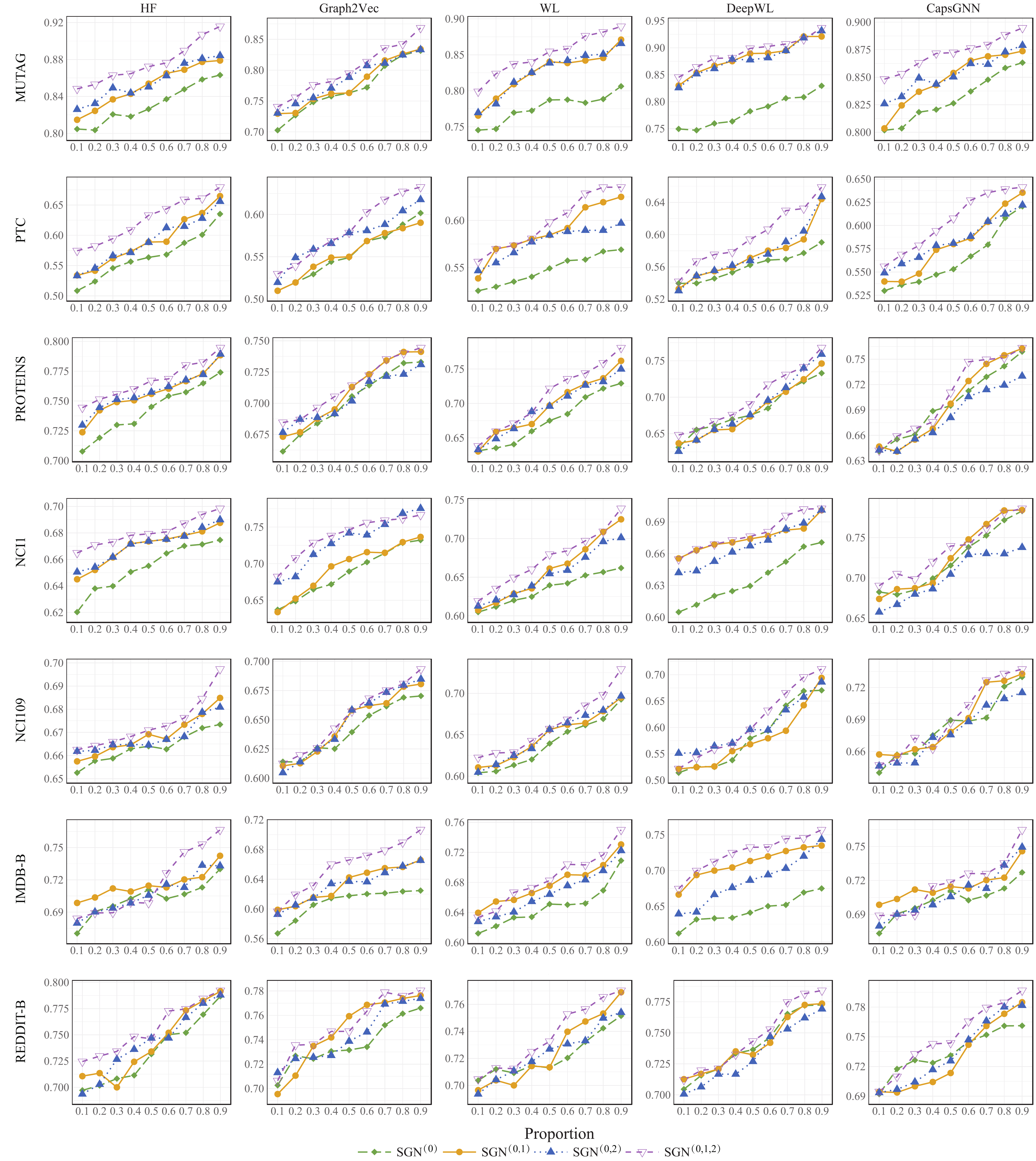}
	\caption{Average $F_1$-$Score$ as functions of the size of the training set (represented by the fraction of samples in the training set), for various feature extraction methods on different datasets, based on SGN$^{\textbf{(0)}}$, SGN$^{\textbf{(0,1)}}$, SGN$^{\textbf{(0,2)}}$ and SGN$^{\textbf{(0,1,2)}}$, respectively.}
	\label{fig:precision}
\end{figure*}

Furthermore, we also visualize the average F1-Scores obtained by using different feature extraction methods under different combinations of SGNs, as shown in Fig.~\ref{fig:exps}. Note that here we also consider the third-order SGNs, in order to present the changing trends of F1-Scores with the number of SGNs more clearly. Indeed, we can find that integrating higher-order SGNs generally helps to capture more structural information, leading to higher classification performance. However, such benefit seems to be shrunk when we go further, i.e., the improvement of F1-Score from SGN$^{\textbf{(0,1,2)}}$ to SGN$^{\textbf{(0,1,2,3)}}$ is relatively small, while the computational complexity increases quite fast. And this is the reason why we only consider first-order and second-order SGNs in most parts of this work.

To address the computational complexity of our method, we record the average execution time to establish SGNs of different orders on the seven datasets, including MUTAG, PTC, PROTEINS, NCI1, NCI109, IMDB-B and REDDIT-B. The results are shown in Fig.~\ref{fig:time}, where we can see that execution time increases fast as the order of SGN and the network size increase. One possible reason is that here the subgraph we chose is relatively simple, making the SGNs of higher-order even more complicated than those of lower-order. Therefore, one way to decrease the computational complexity is to choose more complex subgraphs to establish simpler higher-order SGNs. Another way to accelerate this process is to adopt parallel computing mechanism, which will be our focus in future work.

To address the robustness of the classification model against the size variation of the training set, the $F_1$-$Score$ is calculated for the network classification task, using various sizes of training sets (from 10 to 90 percent, within a 10 percent interval). For each size, the training and test sets are randomly divided, which is repeated for 500 times with the average result recorded. The results are shown in Fig.~\ref{fig:precision} for various feature extraction methods on different datasets. It can be seen that the classification results based on SGN$^{\textbf{(0)}}$, SGN$^{\textbf{(1)}}$ and SGN$^{\textbf{(2)}}$ together are always the best, and the results based on SGN$^{\textbf{(0)}}$ and SGN$^{\textbf{(1)}}$ together, or SGN$^{\textbf{(0)}}$ and SGN$^{\textbf{(2)}}$ together, are always better than those based only on the original network SGN$^{\textbf{(0)}}$. This confirms that the simulation results are quite robust to the variation of the training set size. For further study, our source codes are available online.~\footnote{\url{https://github.com/GalateaWang}}

\section{Conclusions}\label{sec:Con}
In this paper, the concept of subgraph network (SGN) is introduced, along with algorithms developed for constructing the 1st-order and 2nd-order SGNs, which can expand the structural feature space. As a multi-order graph representation method, various orders of SGNs can significantly enrich the structural information and thus benefit the network feature extraction methods to capture various aspects of the network structure. Also, the effectiveness of the 1st-order and 2nd-order SGNs are verified. Moreover, the handcrafted features, as well as the features automatically generated by network representation methods including graph2vec and kernel-based methods including Weisfeiler-Lehman (WL) and deep WL methods and CapsGNN method, are used in experiments for network classification on seven real-world datasets.

The experimental results show that the classification model based on the features of the original network together with the 1st-order and 2nd-order SGNs always performs the best, compared with those based only on a single network, either the original one, the 1st-order or the 2nd-order SGN, or those based on a pair of them. This demonstrates that SGNs can indeed complement the original network on structural information and thus benefit the subsequent network classification algorithms, no matter which feature extraction method is adopted. More interestingly, it is found that the model based on handcrafted features performs even better than those based on the features automatically generated by more advanced methods, such as graph2vec, for most datasets. This finding suggests that, in general, properly chosen structural features with clear physical meanings may be effective in designing structure-based algorithms.

Future research may focus on extracting more types of subgraphs to establish SGNs of higher diversity for both static and temporal networks, so as to capture the network structural information more comprehensively, to design consequent algorithms for network classification and perhaps other tasks as well.

\section*{Acknowledgments}
The authors would like to thank all the members in the IVSN Research Group, Zhejiang University of Technology for the valuable discussion about the ideas and technical details presented in this paper. This work was partially supported by the National Natural Science Foundation of China under Grant 61973273 and Grant 61572439, by the Zhejiang Provincial Natural Science Foundation of China under Grant LR19F030001, and by the Hong Kong Research Grants Council under the GRF Grant CityU11200317.

\ifCLASSOPTIONcaptionsoff
  \newpage
\fi

\bibliographystyle{IEEEtran}
\bibliography{REFs}

\begin{thebibliography}{10}
\providecommand{\url}[1]{#1}
\csname url@samestyle\endcsname
\providecommand{\newblock}{\relax}
\providecommand{\bibinfo}[2]{#2}
\providecommand{\BIBentrySTDinterwordspacing}{\spaceskip=0pt\relax}
\providecommand{\BIBentryALTinterwordstretchfactor}{4}
\providecommand{\BIBentryALTinterwordspacing}{\spaceskip=\fontdimen2\font plus
\BIBentryALTinterwordstretchfactor\fontdimen3\font minus
  \fontdimen4\font\relax}
\providecommand{\BIBforeignlanguage}[2]{{%
\expandafter\ifx\csname l@#1\endcsname\relax
\typeout{** WARNING: IEEEtran.bst: No hyphenation pattern has been}%
\typeout{** loaded for the language `#1'. Using the pattern for}%
\typeout{** the default language instead.}%
\else
\language=\csname l@#1\endcsname
\fi
#2}}
\providecommand{\BIBdecl}{\relax}
\BIBdecl

\bibitem{walter2004visualization}
M.~Walter, C.~Chaban, K.~Sch{\"u}tze, O.~Batistic, K.~Weckermann, C.~N{\"a}ke,
  D.~Blazevic, C.~Grefen, K.~Schumacher, C.~Oecking, K.~Harter, and J.~Kudla,
  ``Visualization of protein interactions in living plant cells using
  bimolecular fluorescence complementation,'' \emph{The Plant Journal},
  vol.~40, no.~3, pp. 428--438, 2004.

\bibitem{wale2008comparison}
N.~Wale, I.~A. Watson, and G.~Karypis, ``Comparison of descriptor spaces for
  chemical compound retrieval and classification,'' \emph{Knowledge and
  Information Systems}, vol.~14, no.~3, pp. 347--375, 2008.

\bibitem{nguyen2018learning}
D.~Nguyen, W.~Luo, T.~D. Nguyen, S.~Venkatesh, and D.~Phung, ``Learning graph
  representation via frequent subgraphs,'' in \emph{Proceedings of the 2018
  SIAM International Conference on Data Mining}.\hskip 1em plus 0.5em minus
  0.4em\relax SIAM, 2018, pp. 306--314.

\bibitem{xuan2018social}
Q.~Xuan, Z.-Y. Zhang, C.~Fu, H.-X. Hu, and V.~Filkov, ``Social synchrony on
  complex networks,'' \emph{IEEE transactions on cybernetics}, vol.~48, no.~5,
  pp. 1420--1431, 2018.

\bibitem{xuan2014focus}
Q.~Xuan, A.~Okano, P.~Devanbu, and V.~Filkov, ``Focus-shifting patterns of oss
  developers and their congruence with call graphs,'' in \emph{Proceedings of
  the 22nd ACM SIGSOFT International Symposium on Foundations of Software
  Engineering}.\hskip 1em plus 0.5em minus 0.4em\relax ACM, 2014, pp. 401--412.

\bibitem{mockus2002two}
A.~Mockus, R.~T. Fielding, and J.~D. Herbsleb, ``Two case studies of open
  source software development: Apache and mozilla,'' \emph{ACM Transactions on
  Software Engineering and Methodology}, vol.~11, no.~3, pp. 309--346, 2002.

\bibitem{kim2018social}
J.~Kim and M.~Hastak, ``Social network analysis: Characteristics of online
  social networks after a disaster,'' \emph{International Journal of
  Information Management}, vol.~38, no.~1, pp. 86--96, 2018.

\bibitem{fu2018link}
C.~Fu, M.~Zhao, L.~Fan, X.~Chen, J.~Chen, Z.~Wu, Y.~Xia, and Q.~Xuan, ``Link
  weight prediction using supervised learning methods and its application to
  yelp layered network,'' \emph{IEEE Transactions on Knowledge and Data
  Engineering}, vol.~30, no.~8, pp. 1507--1518, 2018.

\bibitem{ullmann1976algorithm}
J.~R. Ullmann, ``An algorithm for subgraph isomorphism,'' \emph{Journal of the
  ACM (JACM)}, vol.~23, no.~1, pp. 31--42, 1976.

\bibitem{balazsi2005topological}
G.~Balazsi, A.-L. Barab{\'a}si, and Z.~Oltvai, ``Topological units of
  environmental signal processing in the transcriptional regulatory network of
  escherichia coli,'' \emph{Proceedings of the National Academy of Sciences},
  vol. 102, no.~22, pp. 7841--7846, 2005.

\bibitem{ugander2013subgraph}
J.~Ugander, L.~Backstrom, and J.~Kleinberg, ``Subgraph frequencies: Mapping the
  empirical and extremal geography of large graph collections,'' in
  \emph{Proceedings of the 22nd international conference on World Wide
  Web}.\hskip 1em plus 0.5em minus 0.4em\relax ACM, 2013, pp. 1307--1318.

\bibitem{vohrasubgraph}
\BIBentryALTinterwordspacing
Q.~Vohra, ``Subgraph frequencies and network classification.'' [Online].
  Available:
  \url{http://snap.stanford.edu/class/cs224w-2014/projects2014/cs224w-76-final.pdf}
\BIBentrySTDinterwordspacing

\bibitem{jha2015path}
M.~Jha, C.~Seshadhri, and A.~Pinar, ``Path sampling: A fast and provable method
  for estimating 4-vertex subgraph counts,'' in \emph{Proceedings of the 24th
  International Conference on World Wide Web}.\hskip 1em plus 0.5em minus
  0.4em\relax International World Wide Web Conferences Steering Committee,
  2015, pp. 495--505.

\bibitem{grochow2007network}
J.~A. Grochow and M.~Kellis, ``Network motif discovery using subgraph
  enumeration and symmetry-breaking,'' in \emph{Annual International Conference
  on Research in Computational Molecular Biology}.\hskip 1em plus 0.5em minus
  0.4em\relax Springer, 2007, pp. 92--106.

\bibitem{benson2016higher}
A.~R. Benson, D.~F. Gleich, and J.~Leskovec, ``Higher-order organization of
  complex networks,'' \emph{Science}, vol. 353, no. 6295, pp. 163--166, 2016.

\bibitem{wang2017incremental}
H.~Wang, P.~Zhang, X.~Zhu, I.~W.-H. Tsang, L.~Chen, C.~Zhang, and X.~Wu,
  ``Incremental subgraph feature selection for graph classification,''
  \emph{IEEE Transactions on Knowledge and Data Engineering}, vol.~29, no.~1,
  pp. 128--142, 2017.

\bibitem{yang2018node}
C.~Yang, M.~Liu, V.~W. Zheng, and J.~Han, ``Node, motif and subgraph:
  Leveraging network functional blocks through structural convolution,'' in
  \emph{2018 IEEE/ACM International Conference on Advances in Social Networks
  Analysis and Mining (ASONAM)}.\hskip 1em plus 0.5em minus 0.4em\relax IEEE,
  2018, pp. 47--52.

\bibitem{harary1960some}
F.~Harary and R.~Z. Norman, ``Some properties of line digraphs,''
  \emph{Rendiconti del Circolo Matematico di Palermo}, vol.~9, no.~2, pp.
  161--168, 1960.

\bibitem{thoma2010discriminative}
M.~Thoma, H.~Cheng, A.~Gretton, J.~Han, H.-P. Kriegel, A.~Smola, L.~Song, P.~S.
  Yu, X.~Yan, and K.~M. Borgwardt, ``Discriminative frequent subgraph mining
  with optimality guarantees,'' \emph{Statistical Analysis and Data Mining: The
  ASA Data Science Journal}, vol.~3, no.~5, pp. 302--318, 2010.

\bibitem{wernicke2006efficient}
S.~Wernicke, ``Efficient detection of network motifs,'' \emph{IEEE/ACM
  Transactions on Computational Biology and Bioinformatics (TCBB)}, vol.~3,
  no.~4, pp. 347--359, 2006.

\bibitem{wernicke2005faster}
------, ``A faster algorithm for detecting network motifs,'' in
  \emph{International Workshop on Algorithms in Bioinformatics}.\hskip 1em plus
  0.5em minus 0.4em\relax Springer, 2005, pp. 165--177.

\bibitem{rotabi2017detecting}
R.~Rotabi, K.~Kamath, J.~Kleinberg, and A.~Sharma, ``Detecting strong ties
  using network motifs,'' in \emph{Proceedings of the 26th International
  Conference on World Wide Web Companion}.\hskip 1em plus 0.5em minus
  0.4em\relax International World Wide Web Conferences Steering Committee,
  2017, pp. 983--992.

\bibitem{kovanen2011temporal}
L.~Kovanen, M.~Karsai, K.~Kaski, J.~Kert{\'e}sz, and J.~Saram{\"a}ki,
  ``Temporal motifs in time-dependent networks,'' \emph{Journal of Statistical
  Mechanics: Theory and Experiment}, vol. 2011, no.~11, p. P11005, 2011.

\bibitem{xuan2015temporal}
Q.~Xuan, H.~Fang, C.~Fu, and V.~Filkov, ``Temporal motifs reveal collaboration
  patterns in online task-oriented networks,'' \emph{Physical Review E},
  vol.~91, no.~5, p. 052813, 2015.

\bibitem{paranjape2017motifs}
A.~Paranjape, A.~R. Benson, and J.~Leskovec, ``Motifs in temporal networks,''
  in \emph{Proceedings of the Tenth ACM International Conference on Web Search
  and Data Mining}.\hskip 1em plus 0.5em minus 0.4em\relax ACM, 2017, pp.
  601--610.

\bibitem{tsourakakis2017scalable}
C.~E. Tsourakakis, J.~Pachocki, and M.~Mitzenmacher, ``Scalable motif-aware
  graph clustering,'' in \emph{Proceedings of the 26th International Conference
  on World Wide Web}.\hskip 1em plus 0.5em minus 0.4em\relax International
  World Wide Web Conferences Steering Committee, 2017, pp. 1451--1460.

\bibitem{jing2018deep}
Y.~Jing, Y.~Bian, Z.~Hu, L.~Wang, and X.-Q.~S. Xie, ``Deep learning for drug
  design: An artificial intelligence paradigm for drug discovery in the big
  data era,'' \emph{The AAPS journal}, vol.~20, no.~3, p.~58, 2018.

\bibitem{lane2018comparing}
T.~Lane, D.~P. Russo, K.~M. Zorn, A.~M. Clark, A.~Korotcov, V.~Tkachenko, R.~C.
  Reynolds, A.~L. Perryman, J.~S. Freundlich, and S.~Ekins, ``Comparing and
  validating machine learning models for mycobacterium tuberculosis drug
  discovery,'' \emph{Molecular pharmaceutics}, 2018.

\bibitem{cheng2016wide}
H.-T. Cheng, L.~Koc, J.~Harmsen, T.~Shaked, T.~Chandra, H.~Aradhye,
  G.~Anderson, G.~Corrado, W.~Chai, M.~Ispir \emph{et~al.}, ``Wide \& deep
  learning for recommender systems,'' in \emph{Proceedings of the 1st Workshop
  on Deep Learning for Recommender Systems}.\hskip 1em plus 0.5em minus
  0.4em\relax ACM, 2016, pp. 7--10.

\bibitem{mikolov2013distributed}
T.~Mikolov, I.~Sutskever, K.~Chen, G.~S. Corrado, and J.~Dean, ``Distributed
  representations of words and phrases and their compositionality,'' in
  \emph{Advances in neural information processing systems}, 2013, pp.
  3111--3119.

\bibitem{le2014distributed}
Q.~Le and T.~Mikolov, ``Distributed representations of sentences and
  documents,'' in \emph{International Conference on Machine Learning}, 2014,
  pp. 1188--1196.

\bibitem{narayanan2017graph2vec}
A.~Narayanan, M.~Chandramohan, R.~Venkatesan, L.~Chen, Y.~Liu, and S.~Jaiswal,
  ``graph2vec: Learning distributed representations of graphs,'' \emph{arXiv
  preprint arXiv:1707.05005}, 2017.

\bibitem{vishwanathan2010graph}
S.~V.~N. Vishwanathan, N.~N. Schraudolph, R.~Kondor, and K.~M. Borgwardt,
  ``Graph kernels,'' \emph{Journal of Machine Learning Research}, vol.~11, pp.
  1201--1242, 2010.

\bibitem{shervashidze2011weisfeiler}
N.~Shervashidze, P.~Schweitzer, E.~J.~v. Leeuwen, K.~Mehlhorn, and K.~M.
  Borgwardt, ``Weisfeiler-lehman graph kernels,'' \emph{Journal of Machine
  Learning Research}, vol.~12, pp. 2539--2561, 2011.

\bibitem{yanardag2015deep}
P.~Yanardag and S.~Vishwanathan, ``Deep graph kernels,'' in \emph{Proceedings
  of the 21th ACM SIGKDD International Conference on Knowledge Discovery and
  Data Mining}.\hskip 1em plus 0.5em minus 0.4em\relax ACM, 2015, pp.
  1365--1374.

\bibitem{xuan2018automatic}
Q.~Xuan, B.~Fang, Y.~Liu, J.~Wang, J.~Zhang, Y.~Zheng, and G.~Bao, ``Automatic
  pearl classification machine based on a multistream convolutional neural
  network,'' \emph{IEEE Transactions on Industrial Electronics}, vol.~65,
  no.~8, pp. 6538--6547, 2018.

\bibitem{duvenaud2015convolutional}
D.~K. Duvenaud, D.~Maclaurin, J.~Iparraguirre, R.~Bombarell, T.~Hirzel,
  A.~Aspuru-Guzik, and R.~P. Adams, ``Convolutional networks on graphs for
  learning molecular fingerprints,'' in \emph{Advances in neural information
  processing systems}, 2015, pp. 2224--2232.

\bibitem{bruna2013spectral}
J.~Bruna, W.~Zaremba, A.~Szlam, and Y.~LeCun, ``Spectral networks and locally
  connected networks on graphs,'' \emph{arXiv preprint arXiv:1312.6203}, 2013.

\bibitem{defferrard2016convolutional}
M.~Defferrard, X.~Bresson, and P.~Vandergheynst, ``Convolutional neural
  networks on graphs with fast localized spectral filtering,'' in
  \emph{Advances in Neural Information Processing Systems}, 2016, pp.
  3844--3852.

\bibitem{xinyi2018capsule}
\BIBentryALTinterwordspacing
Z.~Xinyi and L.~Chen, ``Capsule graph neural network,'' in \emph{International
  Conference on Learning Representations}, 2019. [Online]. Available:
  \url{https://openreview.net/forum?id=Byl8BnRcYm}
\BIBentrySTDinterwordspacing

\bibitem{agarwal2006higher}
S.~Agarwal, K.~Branson, and S.~Belongie, ``Higher order learning with graphs,''
  in \emph{Proceedings of the 23rd international conference on Machine
  learning}.\hskip 1em plus 0.5em minus 0.4em\relax ACM, 2006, pp. 17--24.

\bibitem{eckmann2002curvature}
J.-P. Eckmann and E.~Moses, ``Curvature of co-links uncovers hidden thematic
  layers in the world wide web,'' \emph{Proceedings of the national academy of
  sciences}, vol.~99, no.~9, pp. 5825--5829, 2002.

\bibitem{schioberg2015evolution}
D.~Schi{\"o}berg, F.~Schneider, S.~Schmid, S.~Uhlig, and A.~Feldmann,
  ``Evolution of directed triangle motifs in the google+ osn,'' \emph{arXiv
  preprint arXiv:1502.04321}, 2015.

\bibitem{wang2015link}
P.~Wang, B.~Xu, Y.~Wu, and X.~Zhou, ``Link prediction in social networks: the
  state-of-the-art,'' \emph{Science China Information Sciences}, vol.~58,
  no.~1, pp. 1--38, 2015.

\bibitem{li2011graph}
G.~Li, M.~Semerci, B.~Yener, and M.~J. Zaki, ``Graph classification via
  topological and label attributes,'' in \emph{Proceedings of the 9th
  international workshop on mining and learning with graphs (MLG), San Diego,
  USA}, vol.~2, 2011.

\bibitem{xiaofan2012network}
X.~Wang, X.~Li, and G.~Chen, ``Network science: an introduction,'' pp. 87--90,
  2012.

\bibitem{soffer2005network}
S.~N. Soffer and A.~Vazquez, ``Network clustering coefficient without
  degree-correlation biases,'' \emph{Physical Review E}, vol.~71, no.~5, p.
  057101, 2005.

\bibitem{debnath1991structure}
A.~K. Debnath, R.~L. Lopez~de Compadre, G.~Debnath, A.~J. Shusterman, and
  C.~Hansch, ``Structure-activity relationship of mutagenic aromatic and
  heteroaromatic nitro compounds. correlation with molecular orbital energies
  and hydrophobicity,'' \emph{Journal of medicinal chemistry}, vol.~34, no.~2,
  pp. 786--797, 1991.

\bibitem{toivonen2003statistical}
H.~Toivonen, A.~Srinivasan, R.~D. King, S.~Kramer, and C.~Helma, ``Statistical
  evaluation of the predictive toxicology challenge 2000--2001,''
  \emph{Bioinformatics}, vol.~19, no.~10, pp. 1183--1193, 2003.

\bibitem{borgwardt2005protein}
K.~M. Borgwardt, C.~S. Ong, S.~Sch{\"o}nauer, S.~Vishwanathan, A.~J. Smola, and
  H.-P. Kriegel, ``Protein function prediction via graph kernels,''
  \emph{Bioinformatics}, vol.~21, pp. i47--i56, 2005.

\bibitem{joachims2009predicting}
T.~Joachims, T.~Hofmann, Y.~Yue, and C.-N. Yu, ``Predicting structured objects
  with support vector machines,'' \emph{Communications of the ACM}, vol.~52,
  no.~11, pp. 97--104, 2009.

\bibitem{kudo2005application}
T.~Kudo, E.~Maeda, and Y.~Matsumoto, ``An application of boosting to graph
  classification,'' in \emph{Advances in neural information processing
  systems}, 2005, pp. 729--736.

\bibitem{zhao2018substructure}
X.~Zhao, B.~Zong, Z.~Guan, K.~Zhang, and W.~Zhao, ``Substructure assembling
  network for graph classification,'' 2018.

\bibitem{bengio2003neural}
Y.~Bengio, R.~Ducharme, P.~Vincent, and C.~Jauvin, ``A neural probabilistic
  language model,'' \emph{Journal of machine learning research}, vol.~3, pp.
  1137--1155, 2003.

\end{thebibliography}

\end{document}